\documentclass[pra,preprintnumbers,nobalancelastpage,twocolumn,superscriptaddress,showpacs,nofootinbib]{revtex4-1}
\usepackage{amsmath, amssymb}
\usepackage{color}
\usepackage{graphicx}
\usepackage{subfigure}
\usepackage{xifthen}
\usepackage{booktabs}
\usepackage{ulem}
\usepackage[colorlinks=true, citecolor=blue, linkcolor=blue]{hyperref}

\newcommand{\ket}[1]{\ensuremath{\left | #1 \right \rangle}}
\newcommand{\bra}[1]{\ensuremath{\left \langle #1 \right |}}
\newcommand{\tr}[2][]{\ensuremath{\textrm{Tr}_{#1} \left[ #2 \right] }}
\newcommand{\dec}[1][]{
	\ifthenelse{\isempty{#1}}{\ensuremath{\mathcal D_t}}{\ensuremath{\mathcal D_t^{( #1 )} }}
	}
\newcommand{\lpo}[1]{\ensuremath{\sigma^L_{#1} } }
\newcommand{\approxlpo}[1]{\ensuremath{ \widetilde{\sigma}_{#1} } }
\newcommand{\lpos}{\ensuremath{{\boldsymbol \sigma}^L } }
\newcommand{\approxlpos}{\ensuremath{\widetilde{\boldsymbol \sigma} } }
\newcommand{\tracenorm}[1]{\ensuremath{{\| #1 \|}_{\text{tr} }} }

\newcommand{\superket}[1]{\ensuremath{| #1  \rangle\rangle}}
\newcommand{\superbra}[1]{\ensuremath{\langle \langle #1 |}}
\newcommand{\states}[1]{\ensuremath{\mathfrak{S}\left( #1 \right) }}

\newcommand{\fid}[1]{\ensuremath{F_t^{#1}}}

\newcommand{\al}[1]{\begin{align} #1 \end{align}}
\newcommand{\pmat}[1]{\begin{pmatrix} #1 \end{pmatrix} }
\newcommand{\nn}{\nonumber \ }
\newcommand{\eqsys}[1]{\begin{equation} \left\{ \begin{aligned} #1 \end{aligned} \right. \end{equation} }
\newcommand{\set}[1]{\ensuremath{ \left \{ #1 \right\} }}

\newcommand{\mb}[1]{\ensuremath{\mathbf{#1}} }
\newcommand{\mc}[1]{\ensuremath{\mathcal{#1}} }
\newcommand{\bs}[1]{\ensuremath{\boldsymbol{#1}} }

\begin{document}
\def\bbm[#1]{\mbox{\boldmath$#1$}}

\title{A Perturbative Approach to Continuous-Time Quantum Error Correction}

\author{Matteo Ippoliti}
\email{matteoi@princeton.edu}
\address{Scuola Normale Superiore, piazza dei Cavalieri 7,  I-56126 Pisa, Italy}
\address{Department of Physics, Princeton University, Princeton, NJ 08544, USA}

\author{Leonardo Mazza}
\address{NEST, Scuola Normale Superiore and Istituto Nanoscienze-CNR, piazza dei Cavalieri 7, I-56126 Pisa, Italy}

\author{Matteo Rizzi}
\address{Universit\"at Mainz, Institut f\"ur Physik, Staudingerweg 7, D-55099 Mainz, Germany}

\author{Vittorio Giovannetti}
\address{NEST, Scuola Normale Superiore and Istituto Nanoscienze-CNR, piazza dei Cavalieri 7, I-56126 Pisa, Italy}

\begin{abstract}

We present a novel discussion of the continuous-time quantum error correction introduced by Paz and Zurek in 1998 [Paz and Zurek, Proc.~R.~Soc.~A \textbf{454}, 355 (1998)].
We study the general Lindbladian which describes the effects of both noise and error correction in the weak-noise (or strong-correction) regime through a perturbative expansion. 
We use this tool to derive quantitative aspects of the continuous-time dynamics both in general and through two illustrative examples:
the 3-qubit and the 5-qubit stabilizer codes,
which can be independently solved by analytical and numerical methods and then used as benchmarks for the perturbative approach.
The perturbatively accessible time frame features a short initial transient in which error correction is ineffective, followed by a slow decay of the information content consistent with the known facts about discrete-time error correction in the limit of fast operations.
This behavior is explained in the two case studies through a geometric description of the continuous transformation of the state space induced by the combined action of noise and error correction.

\end{abstract}


\maketitle

\section{Introduction \label{sec:intro}}

While the negative effect of an environment on a system
to be used for the storage and manipulation of quantum information
has been often pointed out,
in the last few years it has become clear that decoherence 
can also be a powerful resource~\cite{Verstraete:2009}.
If suitably tamed and engineered,
it can be used to perform several quantum information-processing tasks, including state preparation~\cite{Krauter:2011,  Lin:2013, Bardyn:2013}, 
universal quantum computation~\cite{Verstraete:2009}, 
quantum simulation~\cite{Barreiro:2011, Schindler:2013},
quantum memories~\cite{Pastawski:2011, cellular, Wolf}, 
and quantum control~\cite{qcontrol}.
Although at this point experiments are only at the level 
of proof-of-principle operations,
this new framework provides novel motivation for the revival
and further development of {\it continuous-time} quantum error correction (CTQEC).

CTQEC was originally proposed by Paz and Zurek~\cite{paz-zurek} in 1998 
as a convenient mathematical description
of standard discrete-time quantum error correction (QEC)~\cite{HOLEVOBOOK, Nielsen}
in the limit of frequently repeated operations.
It was later realized that the continuous description
of the error-correcting process could have
a real physical meaning, 
and that it could be realized via continuous quantum feedback control~\cite{Ahn:2002, Oreshkov:2013} or via specific forms of engineered dissipation~\cite{ctqec, sarovar}.

As thoroughly discussed in Ref.~\cite{Wolf}, CTQEC is a specific form of
continuous-time coding scheme, whereby a coding dissipative process competes with a noise and improves the memory performance.
This opens the problem of quantifying the specific properties of these complex dissipative models, e.g.\ how information is corrupted as a function of time, especially when the correction rate is large but finite.

In this article we present an approach to this problem that is based on perturbation theory.
We show how to expand the evolution channel that describes the competing processes of noise and correction as a power series in the noise strength.
This provides a perturbative expansion valid in the strong error correction regime.
By specializing this result to a suitably defined class of ``effective'' quantum recovery operations, we then derive a formula for the approximate channel that describes the loss of information to leading order in the error process.

We test the method on two simple examples based on stabilizer QEC codes~\citep{gottesman}:
the 3-qubit bit-flip code by Shor~\citep{shor} and the 5-qubit perfect code by Laflamme {\it et al.}~\citep{laflamme}.
These codes are simple enough to allow non-perturbative solutions via either analytic or numerical tools.
We derive such solutions and then use them to benchmark our perturbative method quantitatively,
and to gain a qualitative understanding of the key dynamical features that the perturbative approximation tries to capture.
The known results about quantum error correction are recovered in our formalism,
and presented in a novel geometric picture by considering the time evolution as a transformation of the state space;
at the same time, the perturbative approach we describe is shown to successfully reproduce the exact results in a much more efficient and concise way.

The article is organized as follows.
In Sec.~\ref{sec:rec} we review the general concept of recovery operations, and discuss their continuous-time version.
Sec.~\ref{sec:perturb} contains the main result of the article, which is the general perturbative expansion valid in the strong-QEC regime.
The following sections are devoted to the quantitative study of two examples:
in Sec.~\ref{sec:3q} we analyze the 3-qubit bit-flip code, re-deriving the analytical results already obtained by Paz and Zurek~\citep{paz-zurek} through our formalism and presenting a novel discussion of the spectral properties of the Lindbladian;
Sec.~\ref{sec:5q} presents an analogous discussion of the 5-qubit perfect code based on a numerical method.
Both sections are concluded by a quantitative comparison with the perturbative approximation.
Finally, in Sec.~\ref{sec:conclusion} we summarize the results and discuss their relevance.

\section{Continuous-time implementation of QEC recovery operations\label{sec:rec}}

\subsection{Recovery operations}\label{subsec:recovery}

The scenario we are considering is the following. A logical coding Hilbert space ${\cal H}_L$ where quantum information is assumed to be stored, is identified
as a proper subspace 
of an extended physical system (the quantum memory) described by the Hilbert space $\mc H_{\text{memo}}$.
In a general physical situation, the memory is not ideal and errors  may occur from the interaction with the system environment. This will tend to spoil the storing process by introducing decoherence while driving the system outside the coding space. Denoting $\mathfrak{S} (\mathcal H)$ the set of density matrices over the Hilbert space $\mathcal H$, such noise  can be described by assigning a 
completely positive, trace preserving  (CPTP) transformation {$\mc N$}~\cite{HOLEVOBOOK, Nielsen} which maps  elements of  $\states{\mc H_L}$ into those of  $\states{\mc H_{\text{memo}}}$. Specifically if $|\psi_L\rangle \in \mc H_L$ is
the (possibly unknown) initial state of the memory, ${\mc N}[|\psi_L\rangle\langle \psi_L|]$ describes the state of the memory {\it after} the noise has acted.
In this context a recovery operation $\mathcal R : \states{\mc H_{\text{memo}}} \mapsto \states{\mc H_L}$ is a CPTP map which we apply to the corrupted state 
${\mc N}[|\psi_L\rangle\langle \psi_L|]$ with the purpose of restoring the initial state $|\psi_L\rangle$  of the memory. 
Since $\mathcal R$ should act trivially on the logical subspace when no error has occurred,  this transformation is typically assumed to be idempotent, i.e.
\begin{eqnarray} 
\mc R\circ \mc R=\mc R\;,
 \label{idem} \end{eqnarray} 
 where   the symbol $``\circ"$ indicate the composition of super-operators. 
The efficiency of a given $\mc R$ 
   can be quantified by 
comparing the decoded state with the initially encoded one, i.e. by computing
the average of such fidelity over all pure input states $\ket{\psi_L}$~\cite{Mazza:2013}: 
\begin{equation}
F^{\mc R} = \int d \mu_\psi 
\bra{\psi_L} \mc R \circ \mc { \mc N} \left[ \,\ket{\psi_L} \hspace{-0.125cm} \bra{\psi_L} \, \right] \ket{\psi_L}\, ,
\label{eq:average:fidelity}
\end{equation}
where $d\mu_\psi$ is the uniform measure over ${\cal H}_L$. 
The quantity $F^{\mc R}$ gauges how close to its original configuration the state of the embedded qubit can be moved by the operation ${\mc R}$.

Recovery operations are naturally associated with  quantum error correcting procedures. 
Consider for instance a $[n,k]$ stabilizer code where a logical space $\mc H_{L}$ isomorphic to the space $\mc H^{(k)}$
of $k$ logical qubits is encoded into the Hibert space  $\mc H_{\text{memo}}=\mc H^{(n)}$  of $n$ physical qubits by means of $n-k$ {\it stabilizers} $\{ g_i\}$ operating on it
(i.e. $n-k$ mutually commuting Pauli operators of the $n$ physical qubits) \citep{gottesman}.
We remind that in this context $\mc H^{(k)}$ is identified as  
the subspace of states $\ket{\psi}$ such that $g_i\ket{\psi} = \ket{\psi}$ $\forall i$ and that 
the error correction consists in the simultaneous measurement of the stabilizers, which yields a binary string called the {\it syndrome}. A {\it correcting unitary} $U_{\mb s}$, depending on the measured syndrome $\mb s$, is then applied to the code to revert the $\mb s$-syndrome subspace to the code-space via
the condition $U_{\mb s} P_{\mb s} = P U_{\mb s}$~\citep{gottesman}, where 
$P \doteqdot P_{\bf 0} =  \prod_{i = 1}^{n-k}(I+ g_i)/2 $ projects $\mc H^{(n)}$ into  $\mc H_L$,  while
$P_{\mb s} = \prod_{i = 1}^{n-k}(I+(-1)^{s_i} g_i)/2$ are the \textit{syndrome projectors} of the code.
Such two-stage, measure-and-correct procedure can be easily cast into the form of a single CPTP recovery operation 
 by introducing the mapping 
\begin{align}
\mc R [\rho] 
& = \sum_{\mb s}  U_{\mb s} P_{\mb s} \rho P_{\mb s} U_{\mb s}^\dagger \, ,
\label{eq:stab-rec0} \\
& = P \left( \sum_{\mb s}  U_{\mb s} \rho U_{\mb s}^\dagger \right) P \,,
\label{eq:stab-rec}
\end{align}
the trace preservation  being ensured by the completeness relation $\sum_{\mb s} (P_{\mb s} U_{\mb s}^\dagger) (U_{\mb s} P_{\mb s})  =\sum_{\mb s} P_{\mb s} =  I$, while the
idempotent property~(\ref{idem}) by the orthogonality of the projectors~$P_{\mb s}$. 

\subsection{Continuous-time implementation}

In the previous section we assumed that the recovery transformation $\mc R$ acts {\it after} the noise $ \mc N$ has affected the memory system.
In what follows instead we are interested in the case where {\it both} the noise and the associate recovery transformation operate continuously over time on the system. 
In particular we will assume the former to be generated  by a Markov process 
$\mathcal N_t = e^{\eta t \mc L_{\text{n.}}}$
induced by a Lindblad super-operator  $\mc L_{\text{n.}}$, while a recovering QEC map~(\ref{eq:stab-rec}) is applied stochastically with a probability of success which, over a time interval $\Delta t$, increases exponential with a characteristic time-scale $\gamma^{-1}$, i.e. 
\begin{equation}
\rho \stackrel{\text{recovery}}{\longrightarrow} e^{-\gamma \Delta t}\rho + \left(1-e^{-\gamma \Delta t} \right) \mc R \left[\rho\right] \,.
\label{rhooft}
\end{equation}
Following~Ref.~\cite{paz-zurek} we model this scenario by describing the dynamical evolution of the system in terms of the master equation obtained by adding to the
noise generator $\mc L_{\text{n.}}$ a contribution that originates from the process~(\ref{rhooft}), i.e. 
\begin{eqnarray}
\frac{d}{dt} \rho(t)  &=& \eta \mc L_{\text{n.}}[\rho(t)] + \gamma \mc L_{\text{e.c.}}[\rho(t)]\;,
\label{eq:mastereq111} 
\end{eqnarray}
where the rate $\eta$ defines the characteristic time-scale of the noise and where 
\begin{equation}
 \mc L_{\text{e.c.}}[\cdots] \doteqdot \mc R [\cdots] - \mc I[\cdots]\,,
\label{eq:mastereq0}
\end{equation}
the symbol $\mc I[\cdots]$ representing the identity channel.
 It goes without mentioning that typically
 the dissipative process described by $\mc L_{\text{e.c.}}$ does not arise naturally, and has to be carefully engineered.
In Ref.~\cite{Verstraete:2009, Pastawski:2011} 
a general way to engineer arbitrary forms of Markovian dissipation was shown, which employs both Hamiltonian interactions and damped qubits: as one ancilla per Lindblad operator is needed, the recovery operator associated with a $[n,k]$ stabilizer code requires a total of $2^{n-k}$ ancillas; rigorous error estimates are also provided.

The dynamics described by~(\ref{eq:mastereq111}) is governed by two independent mechanisms which compete with each other. 
On the one hand, the noise associated with the 
 generator $\mc L_{\text{n.}}$, in the absence of $\mc L_{\text{e.c.}}$, will tend  to corrupt the coded state by ``moving'' it out of the coding space $\mc H_{L}$.
 On the other hand the generator  $\mc L_{\text{e.c.}}$ tends instead to ``freeze" the dynamics induced by the noise by forcing the system to remain in  $\mc H_{L}$.
 The efficiency of  $\mc L_{\text{e.c.}}$ can be measured again through the average fidelity~\eqref{eq:average:fidelity}. More specifically,  we will adopt the following figure of merit:
\begin{eqnarray}
\fid{\mc T} &=& \int d \mu_\psi 
\bra{\psi_L}  \mc T \circ {\dec[\eta] } \left[ \,\ket{\psi_L} \hspace{-0.125cm} \bra{\psi_L} \, \right] \ket{\psi_L}\, ,
\label{eq:average:fidelity1}
\end{eqnarray}
where $\mc T$ is a generic CPTP map and 
\begin{eqnarray}  \label{formalsol} 
\dec[\eta] = \exp \left[ t \left( \eta \mc L_{\text{n.}} + \gamma \mc L_{\text{e.c.}} \right)\right]\;,
\end{eqnarray}
is the CPTP mapping that solves Eq.~(\ref{eq:mastereq111}) -- the dependence on the parameter $\eta$ is emphasized
for future reference.
In the special $\mc T = \mc I$ case, Eq.~(\ref{eq:average:fidelity1}) yields  the bare average fidelity between the input states of the logical space  and their evolved counterparts under the action of the
noise and of the instantaneous error corrections, as a function of the elapsed time $t$. 
The quantity $\fid{\mc T}$, with $\mc T \neq \mc I$, measures the same average fidelity under the assumption that an additional, finite, non-trivial recovery operation $\mc T$ (possibly $\mc R$ itself)  is performed at the read-out.
 Clearly if $\mc T$ provides a reasonable protection for the selected noise, then $\fid{\mc T}$ should be always larger than $\fid{\mc I}$ as it will remove all the spurious 
errors left in the system by the instantaneous, but weak, QEC applications.
The figure \fid{\mc I} may nonetheless be of interest in all cases in which performing a non-trivial recovery at read-out (as opposed to a simple projective measurement) is technically complicated.

%----------------------------------------------------------------------------------------
%----------------------------------------------------------------------------------------
% PERTURBATION THEORY
%----------------------------------------------------------------------------------------
%----------------------------------------------------------------------------------------

\section{Perturbative expansion \label{sec:perturb}}

In this Section we present the main result of the article.
We shall consider the evolution channel $\dec[\eta]$ in~(\ref{formalsol}) and study it in the limit where the noise contribution associated with $\mc L_{\text{n.}}$ is weaker than the one associated with
the error correction contribution~$\mc L_{\text{e.c.}}$.
Formally this is obtained by assuming $\eta$ to be the smallest rate of the system (i.e. $\eta \ll \gamma, 1/t$) and  expanding $\dec[\eta]$ perturbatively with respect  to such parameter. 
For the moment no extra assumptions will be made either on the specific form of   $\mc L_{\text{n.}}$, nor on the recovering map  $\mc R$ which defines  $\mc L_{\text{e.c.}}$  --  apart from the idempotent relation~(\ref{idem}) which we always take for granted~\cite{NOTA}.
 %the time-ordered exponent is given by 
% \begin{widetext}
%  \begin{eqnarray} \label{series} 
% \overleftarrow{\exp}[\int_0^t dt'  \tilde{\mc L}_{\text{n.}}(t') ] =  \mc I +  \sum_{k=1}^\infty \eta^k    \int_0^t dt_{1} \int_0^{t_1} dt_{2} \cdots \int_0^{t_{k-1}}dt_{k}\;\;
%  \tilde{\mc L}_{\text{n.}}(t_1)\circ\tilde{\mc L}_{\text{n.}}(t_2)\circ \cdots \circ  \tilde{\mc L}_{\text{n.}}(t_k)\;.
% \end{eqnarray}
% \end{widetext}  

 In what follows we shall limit the analysis to study the dynamics induced by $\mc D_t^{(\eta)}$ on logical states, i.e. inputs $\rho$ which have support on the codomain 
 $\states{\mc H_L}$
of $\mc R$. This amounts to focusing on the channel 
\begin{eqnarray} \label{dlog}
\mc D_t^{(\eta)}\circ \mc R  = \mbox{CTQEC on logical states,} 
\end{eqnarray} 
and to compare its performance with effect of a bare noisy evolution,
i.e. 
\begin{eqnarray} \label{nolog}
 \mc N_t \circ \mc R  = \mbox{NOISE on logical states.}
\end{eqnarray} 

Along the line detailed when introducing  Eq.~(\ref{eq:average:fidelity1}), 
both regimes will be also studied by including a final recovery  operation $\mc T = \mc R$ acting on the system at the end of 
the dynamical process, i.e. studying the performances of 
\begin{eqnarray} \label{rdlog}
\mc R \circ \mc D_t^{(\eta)}\circ \mc R = \begin{array}{c} \mbox{\small CTQEC on logical states } \\\mbox{\small with final recovery operation,} \end{array} 
\end{eqnarray} 
and 
\begin{eqnarray} \label{rnolog} 
\mc R \circ  \mc N_t   \circ \mc R = \begin{array}{c} \mbox{\small NOISE on logical states} \\ \mbox{\small with final recovery operation.} \end{array} 
\end{eqnarray}   
The last mapping in particular can be used to induce a classification for the recovery operations $\mc R$. For this purpose we expand $\mc N_t$ as a power series
 in terms of its generator, i.e.
\begin{eqnarray} 
\mc R \circ  \mc N_t   \circ \mc R  = \mc R + \sum_{k=1}^\infty \; \frac{(\eta t)^k}{k!} \; \mc R \circ \mc  L_{\text{n.}}^k \circ \mc R \;,
\end{eqnarray} 
with  $\mc  L_{\text{n.}}^k$ indicating $k$ applications of $\mc  L_{\text{n.}}$, e.g. $ \mc  L^2_{\text{n.}}= \mc  L_{\text{n.}}\circ  \mc  L_{\text{n.}}$.
A map $\mc R$ is then said to be $k$-effective for the noise $\mc N_t$  if for all integers $k' \in \{1,\dots k\}$ we have 
\begin{eqnarray} \label{keff}
 \mc R \circ \mc  L_{\text{n.}}^{k'} \circ \mc R = 0\;,
 \end{eqnarray} 
the rationale being that under this condition  it takes at least $k+1$ iterations of the noise generator to move the system out of the coding space.
 Physically, that means we require the code to successfully correct {\it any sequence of $k$ errors}; for quantum codes that deal with single-qubit errors, this can be phrased in terms of the familiar concept of distance \citep{gottesman}.
We emphasize the fact that the notion of $k$-effectiveness is strictly dependent upon the given noise Lindbladian, and that therefore we are in general discussing a form of {\it channel-adapted} quantum error recovery~\citep{fletcher-phd, fletcher}.
We will thus say that a CTQEC scheme $\mc R$ is effective against a noise $\mc L_{\text{n.}}$ if it successfully cancels the first order, i.e. if it is 1-effective. 
Canceling higher orders as well is a desirable property, but it is not necessary.

\subsection{First-order expansion}

For $\eta=0$ Eq.~(\ref{formalsol}) yields the channel
 \begin{eqnarray}
 \dec[0] = e^{\gamma t \mc L_{\text{e.c.}}} = e^{-\gamma t} \mc I + (1-e^{-\gamma t})\mc R \;,
 \end{eqnarray} 
which is responsible for  the transformation~(\ref{rhooft}) and which, in our perturbative  approach,  represents the free evolution of the system.
 For future reference we notice that from the  idempotence property \eqref{idem} it follows that 
 \begin{eqnarray} \label{relaz}
 \mc R \circ  \dec[0] =  \dec[0] \circ \mc R= \mc R\;.
 \end{eqnarray} 
For $\eta\neq 0$ an explicit expression for $\dec[\eta]$ is obtained in terms of the following  Dyson expansion series, 
\begin{eqnarray} \label{EXACT}
 \dec[\eta] =  \dec[0]  \circ \overleftarrow{\exp}\Big[  \int_0^t dt'  \; \eta \, \tilde{\mc L}_{\text{n.}}(t') \Big]\;,
\end{eqnarray} 
where $\tilde{\mc L}_{\text{n.}}(t)$ is the  noise super-operator expressed in the interaction picture induced by $\mc D_t^{(0)}$, i.e. 
\begin{eqnarray} 
 \tilde{\mc L}_{\text{n.}}(t)  = \mc D_{-t}^{(0)} \circ {\mc L}_{\text{n.}}\circ \mc D_{t}^{(0)}\;,
 \end{eqnarray} 
and where
$\overleftarrow{\exp}$ denotes the time-ordered exponential.

The first-order correction in the expansion parameter $\eta$ of~(\ref{EXACT}) is obtained by truncating the Dyson series at $k=1$.
Exploiting~(\ref{relaz})  this gives  
% \begin{widetext}
\begin{align}
\mc D_t^{(\eta)} \circ \mc R 
& = \mc R + \eta  \int_0^t dt_1\;  \mc D_{t-t_1}^{(0)}  \circ \mc L_{\text{n.}} \circ  \mc R  + \mc O(\eta^2) \nn \\
& =	\begin{aligned}[t] 
	& \mc R + \frac{\eta}{\gamma} (1- e^{-\gamma t}) \; \mc L_{\text{n.}} \circ \mc R  \\
	& -  \frac{\eta}{\gamma}  (1- e^{-\gamma t} - \gamma t) \; \mc R\circ   \mc L_{\text{n.}} \circ \mc R + \mc O(\eta^2)\;,
	\end{aligned}
\label{EQNEW}
\end{align} 
%\end{widetext} 
for the CTQEC process~(\ref{dlog}), and 
\begin{eqnarray} \label{EQNEW0}
\mc R \circ \mc D_t^{(\eta)} \circ \mc R 
= \mc R +  \eta t \;  \mc R\circ  \mc L_{\text{n.}} \circ \mc R  + O(\eta^2)\;,
\end{eqnarray}
for the CTQEC process followed by a finite recovery operation $\mc R$ at the end of the dynamical evolution~(\ref{rdlog}).
Notice that these expressions makes it explicit that the expansion we are performing can be trusted only for $\eta\ll \gamma$ and  up to  times $t \ll 1/\eta$. 

In the absence of  CTQEC instead we get 
\begin{eqnarray} \label{EQNEW111}
 \mc N_t   \circ \mc R 
 &=& \mc R + \eta t \;  \mc L_{\text{n.}} \circ \mc R + \mc O(\eta^2) \;, \\
\mc R\circ   \mc N_t   \circ \mc R \label{EQNEW112} 
 &=& \mc R + \eta t \; \mc R \circ  \mc L_{\text{n.}} \circ \mc R + \mc O(\eta^2) \;,
\end{eqnarray} 
for (\ref{nolog}) and (\ref{rnolog}) respectively. 

\begin{figure}[b]
\centering
\includegraphics[width=0.9\columnwidth]{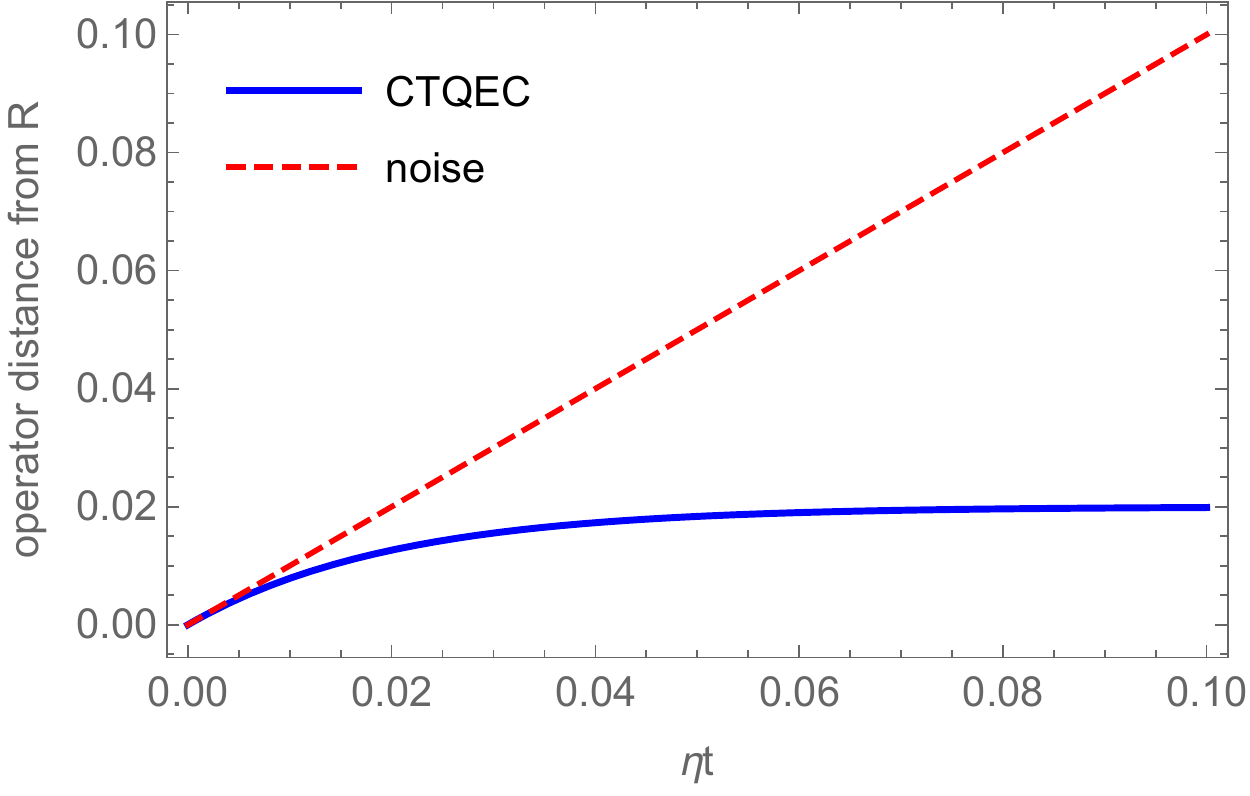}
\caption{
(Color online) Normalized operator distance between the first-order approximation of $\dec[\eta]\circ \mc R$ [Eq.~\eqref{EQNEWNEW}] and $\mc R$ as a function of time (solid line);
same for ${\cal N}_t \circ \mc R$ [Eq.~\eqref{EQNEW111}] (dashed line). 
The distances are normalized by $\|\mc L_{\text{n.}} \circ \mc R \|$,
so that the result does not depend on the choice of operator norm.
 \label{fig:opdistance1}}
\end{figure}

Consider hence the case where the recovery transformation $\mc R$ is  at least a $1$-effective  ~(\ref{keff})  with respect to the noise $\mc N_t$, i.e.   $\mc R \circ \mc L_{\text{n.}} \circ \mc R =0$.
In this limit Eq.~(\ref{EQNEW}) becomes 
\begin{eqnarray} \label{EQNEWNEW}
\mc D_t^{(\eta)}\circ \mc R = \mc R + \frac{\eta}{\gamma} (1- e^{-\gamma t}) \; \mc L_{\text{n.}} \circ \mc R  + \mc O(\eta^2)\;,
\end{eqnarray} 
which by direct comparison with its purely noisy counterpart (\ref{EQNEW111}) shows a clear improvement in the system performance: the linear departure from $\mc R$ (the identity transformation on the coding  space) is in fact now replaced by an asymptotic constant term which for $t \gg \gamma^{-1}$ is proportional to $\eta /\gamma$ (see Fig.~\ref{fig:opdistance1}).

The advantage gained by exploiting the CTQEC  procedure however seems to be washed away when we include a single, finite recovery transformation acting on the system
at the very end of the dynamical process. 
In this case in fact, from Eq.~(\ref{EQNEW0}) and (\ref{EQNEW112}) it follows that for a $1$-effective recovery transformation $\mc R$, 
 both $\mc R \circ \mc D_t^{(\eta)} \circ \mc R$ and $\mc R\circ   \mc N_t   \circ \mc R$ coincide with $\mc R$ up to corrections of order $\eta^2$: 
 to evaluate the effects of CTQEC in this regime one needs to include higher expansion terms, a task that we accomplish in the next subsection.

\subsection{Second-order expansion \label{sec:perturb2nd}}

Second-order effects in $\eta$ can be included by truncating the Dyson series at $k=2$. 
We report here the results  for $1$-effective maps $\mc R$ which are the only ones which are relevant for our purposes (if the QEC procedure is not able to
correct at least first order effects it is useless to analyze second order effects). 
In this regime we get (see Appendix~\ref{app:perturb} for additional details):
 \begin{widetext}
\begin{eqnarray}\label{EQNEW1221}
\mc D_t^{(\eta)} \circ \mc R 
& =& \mc R + \eta  \int_0^t dt_1 \;  \mc D_{t-t_1}^{(0)}  \circ \mc L_{\text{n.}} \circ  \mc R  + {\eta^2}   \int_0^t dt_1 \int_0^{t_1} dt_2 \;  \mc D_{t-t_1}^{(0)}  \circ \mc L_{\text{n.}} \circ
\mc D_{t_1-t_2}^{(0)}  \circ  \mc L_{\text{n.}}  \circ \mc R + \mc O(\eta^3)  \\
&=& \mc R + \tfrac{\eta}{\gamma} (1- e^{-\gamma t}) \; \mc L_{\text{n.}} \circ \mc R   + \tfrac{\eta^2}{\gamma^2} \left\{ [1 - e^{-\gamma t} (1 + \gamma t) ]\;
\mc I + [\gamma t -1 +(\gamma t+2) e^{-\gamma t}  ]
\mc R \right\} \circ \mc L_{\text{n.}}^2 \circ \mc R  \; 
 + \mc O(\eta^3) \;. \nn
\end{eqnarray} 
\end{widetext} 
an expression which simplifies further when including also a finite recovery at time $t$, i.e. 
\begin{eqnarray}
\mc R \circ \mc D_t^{(\eta)} \circ \mc R 
&=& \mc R  + \tfrac{\eta^2}{\gamma^2}\; [\gamma t - 1+ e^{-\gamma t}  ]\;
\mc R\circ \mc L_{\text{n.}}^2 \circ \mc R 
 \nonumber\\ 
&&  \qquad \qquad \qquad \qquad + \mc O(\eta^3) \label{EQNEW1222}\;.
\end{eqnarray} 

\begin{figure}
\centering
\includegraphics[width=0.9\columnwidth]{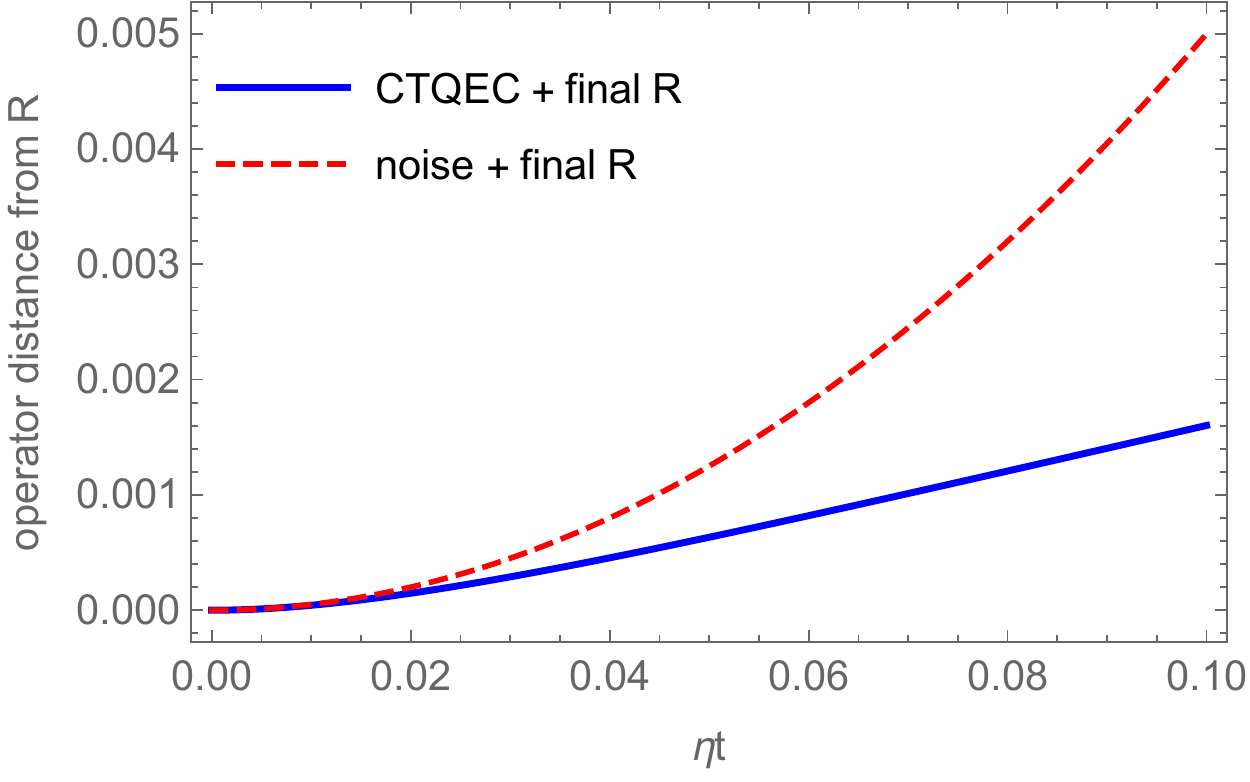}
\caption{
(Color online) Normalized operator distance between the second-order approximation of $\mc R \circ \dec[\eta]\circ \mc R$ [Eq.~\eqref{due2}] and $\mc R$ as a function of time (solid line);
same for $\mc R \circ {\cal N}_t \circ \mc R$ [Eq.~\eqref{ff}] (dashed line). 
The distances are normalized by $\|\mc R \circ \mc L_{\text{n.}}^2 \circ \mc R \|$,
so that the result does not depend on the choice of operator norm.
 \label{fig:opdistance2}}
\end{figure}

From \eqref{EQNEW1222} we can gain some insight in the real-time dynamics of an effective CTQEC scheme.
In particular we noticed that for an initial transient defined by the condition $t \ll 1/\gamma$, the system exhibits a quadratic  dependence  on $t$, i.e. 
\begin{eqnarray}  \label{uno1} 
\mc R \circ  \mc D^{(\eta)}_t \circ \mc R \approx \mc R +\frac{1}{2}(\eta t)^2 \mc R \circ \mc L^2_{\text{n.}}  \circ \mc R \;,
\end{eqnarray} 
which exactly  coincides with the one we would get from~(\ref{rnolog}) when studying the evolution of the system in the {\it absence}  of the CTQEC for a $1$-effective map 
in the presence of a final recovery operation, i.e. 
\begin{eqnarray}  \label{ff} 
\mc R \circ  \mc N_t   \circ \mc R = \mc R + \frac{1}{2}  {(\eta t)^2} \; \mc R\circ \mc L_{\text{n.}}^2 \circ \mc R + \mc O(\eta^3)\;. 
\end{eqnarray} 
At later times $t\gg 1/\gamma$ (but still $t\ll 1/\eta$) Eq.~(\ref{uno1}) however gets replaced by a linear scaling 
\begin{eqnarray} \label{due2} 
\mc R \circ \mc D^{(\eta)}_t \circ \mc R \approx \mc R + \frac{\eta^2 t}{\gamma} \; \mc R \circ \mc L^2_{\text{n.}}   \circ \mc R\;, 
\end{eqnarray} 
which clearly outperforms  Eq.~(\ref{ff})  indicating that the error correction mechanisms is becoming more effective as $t$ increases -- see Fig.~\ref{fig:opdistance2}. 
This behavior agrees qualitatively with what we should expect from the familiar ``discrete-time'' version of QEC (which must be well-approximated by CTQEC in the fast-operation limit):
if the correction operation $\mc R$ is iterated with an average time interval $\tau\simeq 1/\gamma$ between consecutive applications
 the first error correction operation is expected to occur at $t\gtrsim \tau$. After this point, fidelity losses are entirely due to undetectable logical errors, which only occur when 2 elementary errors happen in the time $\tau$ between consecutive QEC operation. The probability for such an event is $\sim (\eta \tau)^2$, hence the rate $\eta^2 \tau \simeq \eta^2 / \gamma$.

From this discussion one expects that if $\mc R$ is $k$-effective, with $k\geq 2$, then fidelity losses should be further suppressed by factors of $\eta/\gamma$. Indeed, one can easily show (see Appendix~\ref{app:keffect}) that for a $k$-effective $\mc R$ the leading correction to the trivial evolution $\mc R$ is
\begin{align}
\mc R \circ \dec[\eta] \circ \mc R -\mc R 
& \approx (\eta/\gamma)^{k+1} f_k (\gamma t) \mc R \circ \mc L_{\text{n.}}^{k+1} \circ \mc R \;,
\label{eq:k-eff-1}
\end{align}
where
\begin{equation}
f_k(x) = x - \int_0^x dy\;   e^{-y} \, \sum_{l=0}^{k-1} \frac{1}{l!} y^l \; .
\label{eq:k-eff-2}
\end{equation}
We observe that for $x\ll 1$ one has $f_k(x)\sim x^{k+1}$, therefore the short-time behavior of the leading-order correction \eqref{eq:k-eff-1} is $\sim (\eta t)^{\kappa+1}$, which is again insensitive to $\gamma$. Then this is smoothly matched to the long-time decay rate $\eta^{k+1}/\gamma^k$.

In the remainder of the paper we shall however focus on the most practically relevant case of 1-effective maps.

\subsection{Stabilizer quantum codes}

As an application of the previous arguments we now specialize  to the case of a $[n,k]$ stabilizer quantum code. Consider hence 
the noise Lindbladian
$\mc L_{\text{n.}}(\rho) = \sum_{\mb s \in \mathfrak E} (U_{\mb s} \rho U_{\mb s}^\dagger - \rho)$, where the $U_{\mb s}$ are Pauli operators defined on the space of $n$ physical qubits and $\mb s$ labels the syndrome. $\mb s$ runs over a subset $\mathfrak E$ of the allowed $2^{n-k}$ syndrome strings; for some codes and noise models (such as the instances we analyze in the following sections \ref{sec:3q} and \ref{sec:5q}), this subset is merely $\{\mb s \neq \bs 0\}$, while in general (e.g. for the Shor code under general single-qubit noise, in which case one has $2^8$ possible syndromes only 27 of which are effectively used to correct errors) $\mathfrak E$ is a smaller subset. 
The only assumption we make is that different errors yield different syndromes. For this model  a $1$-effective   recovery transformation is provided by the mapping~(\ref{eq:stab-rec0}) where, at variance with $\mc L_{\text{n.}}$, the sum over $\mb s$ is not restricted to just $\mathfrak E$ but includes all possible $2^{n-k}$  binary strings.
To see this explicitly, let $N=|\mathfrak E|$ and assume $\rho$ to be a logical state, $\mc R(\rho) = \rho$. Then we can write 
\begin{align}
&   \mc R \circ \mc L_{\text{n.}} \circ \mc R(\rho)
= \sum_{\mb p, \mb q} \sum_{\mb s \in \mathfrak E} P U_{\mb p} U_{\mb s} P U_{\mb q} \rho U^\dagger_{\mb q} P U_{\mb s}^\dagger U_{\mb p}^\dagger P - N \rho \nn \\
& \qquad = \sum_{\mb p, \mb q} \sum_{\mb s \in \mathfrak E} U_{\mb p} U_{\mb s} P_{\mb s \oplus \mb p }P U_{\mb q} \rho U^\dagger_{\mb q} P P_{\mb s \oplus \mb p } U_{\mb s}^\dagger U_{\mb p}^\dagger - N \mc \rho \nn \\
& \qquad = \sum_{\mb q} \sum_{\mb s \in \mathfrak E} P U_{\mb q} \rho U_{\mb q}^\dagger P - N \mc R(\rho)  = 0\,,
\label{eq:stab}
\end{align}
where $\mb s \oplus \mb p = \mod(\mb{s+b},2)$ and
we used the fact that $ P_{\mb s \oplus \mb p }P=0$ unless $\mb s = \mb p$ and that $(U_{\mb s})^2 = 1$, being a Pauli operator.
Accordingly for this model we can use Eq.~(\ref{EQNEW1222}) to estimate 
 the rate of information loss in the presence of the  corresponding CTQEC followed by a final recovery.
We can simplify this process by introducing the super-operator $\mc C$ defined by $\mc L_{\text{n.}} = \mc C - N \mc I$ and observing that  the perturbative 
term on the right-hand-side of Eq.~(\ref{EQNEW1222}) rewrites as 
\al{
\mc R \circ \mc L_{\text{n.}}^2 \circ \mc R
& = \mc R \circ \mc C \circ (\mc C - \mc I) \circ \mc R \nn \\ 
& =  \mc R \circ \mc C^2 \circ \mc R - N \mc R \circ \mc C \circ \mc R \nn \\
& = \mc R \circ \mc C^2 \circ \mc R - N^2 \mc R\,,
}
where in the last passage we used the fact that $\mc R \circ \mc C \circ \mc R = N \mc R$ which trivially follows from Eq.~(\ref{eq:stab}). 
Thus the second-order correction is known once we are able to compute
\al{
\mc R \circ \mc C^2 \circ \mc R (\rho)
& = \sum_{\mb p} \sum_{\mb r, \mb s \in \mathfrak E}P U_{\mb p} U_{\mb r} U_{\mb s} \rho U_{\mb s}^\dagger U_{\mb r}^\dagger U_{\mb p}^\dagger P \nn \\
& = \sum_{\mb p} \sum_{\mb r, \mb s \in \mathfrak E} U_{\mb p} U_{\mb r} U_{\mb s} P_{\mb p \oplus \mb r \oplus \mb s} \rho (\cdots)^\dagger \,,
}
where $(\cdots)^\dagger$ denotes the adjoint of the operator acting on the left.
Now, since $\rho = P \rho P$, the summand is non-zero only if $\mb p = \mb r \oplus \mb s$ ($\mb p$ ranges over all syndromes so this non-zero term is always present), so we have
\al{
\mc R \circ \mc L_{\text{n.}}^2 \circ \mc R (\rho)
& = \sum_{\mb r, \mb s \in \mathfrak E} ( U_{\mb r \oplus \mb s} U_{\mb r} U_{\mb s} \rho U_{\mb s}^\dagger U_{\mb r}^\dagger U_{\mb r \oplus \mb s}^\dagger  - \rho ).
\label{eq:2nd-order-stab}
}
The summand is non-zero whenever a syndrome $\mb r \oplus \mb s$ coming from two distinct errors is corrected through a single unitary that does not undo the effect of the two errors. 
We thus recover the familiar discrete-time picture, in which the fidelity loss is due to multiple errors being misinterpreted as a single one and hence turned into an undetectable logical error by the ``correction'' process.

%----------------------------------------------------------------------------------------
%----------------------------------------------------------------------------------------
% CTQEC ON 3 QUBITS
%----------------------------------------------------------------------------------------
%----------------------------------------------------------------------------------------

\section{CTQEC on the 3-Qubit Bit-Flip Code\label{sec:3q}}

We now focus on the simplest QEC code, i.e.\ the 3-qubit bit-flip code.
This code is simple enough to allow an exact analytical solution for the continuos-time implementation,
which is ideal for investigating in greater detail some of the general features which we demonstrated in the previous Section.

\subsection{The Model}

The 3-qubit QECC is a stabilizer code defined by two independent stabilizers, $g_1=Z_1 Z_2$ and $g_2=Z_2 Z_3$ (here and in the following, $X_{i}$, $Y_{i}$, $Z_{i}$ are the Pauli operators acting on the $i$-th real qubit). The code-space projector is $P = \prod_{i = 1}^2 (1+g_i)/2= \frac{1}{4}(I+Z_1Z_2)(I+Z_2Z_3)$. The logical operators are denoted $\bar X $, $\bar Y$ and $\bar Z$, where $\bar X =  X_1 X_2 X_3$ and similarly for the other two. Note that logical operators commute with $P$, as well as with the projectors $P_{\bf s} = \frac 14 (I+ (-1)^{s_1} Z_1Z_2)(I+ (-1)^{s_2} Z_2Z_3)$, where $\mathbf s = (s_1, s_2)^T$ and $s_i \in \{ 0,1 \}$.
It follows from~\eqref{eq:mastereq0} that the error-correcting Lindblad operators are $\{P ,  P X_1, P X_2, PX_3\}$, with an intensity coefficient $\gamma$.

As for the noise, we consider a bit-flip noise acting identically and independently on each qubit with decay rate $\kappa$:
\begin{align}
 \mc N_t
& = \phi_t^{\otimes 3}, & 
\phi_t (\sigma_\alpha) 
& =
	\begin{cases}
	\sigma_{\alpha}, & \alpha=0,1\;, \\
	e^{-\kappa t} \sigma_{\alpha}, & \alpha=2,3\;,
	\end{cases}
\label{eq:3q-noise}
\end{align}
which can be represented by a generator $\mc L_{\text{n.}}$ characterized by  Lindblad operators 
$\{ X_1,   X_2, X_3\}$ and by  an intensity coefficient $\eta = \frac{\kappa}{2}$.
Accordingly, adopting $\mc R$ as in Eq.~(\ref{eq:stab-rec0}),  Eq.~(\ref{eq:mastereq111}) yields
\begin{align}
\frac{d}{dt} \rho 
 =& \eta \mc L_{\text{n.}}[\rho] + \gamma \mc L_{\text{e.c.}}[\rho] 
 = \frac{\kappa}{2} \left( \sum_{i=1}^3 X_i \rho X_i - 3 \rho \right) + \nonumber \\
&
+\gamma \left( P  \rho P + \sum_{i=1}^3 PX_i  \rho X_i P 
 -  \rho \right) .
\label{eq:3q-ctqec-mastereq-final}
\end{align}

\subsection{Solution of the dynamics}

An analytical solution for Eq.~(\ref{eq:3q-ctqec-mastereq-final}) can be found by observing that both the initial condition $\rho(0) = P \rho(0) P$ and the evolution equation are invariant under the exchange of any two qubits.
This symmetry justifies the following ansatz:
\begin{align}
\rho(0) 
\xrightarrow{ \dec[\eta] } \rho(t) \doteqdot & \; a(t) \rho(0) + b(t) \sum_{i=1}^3 X_i \rho(0) X_i + \nonumber \\
&+c(t) \sum_{i=1}^3 X_i \bar X \rho(0) \bar X X_i 
+ d(t) \bar X \rho(0) \bar X\,.
\label{eq:3q-ansatz}
\end{align}

\begin{figure}
	\centering
	\includegraphics[width=\columnwidth]{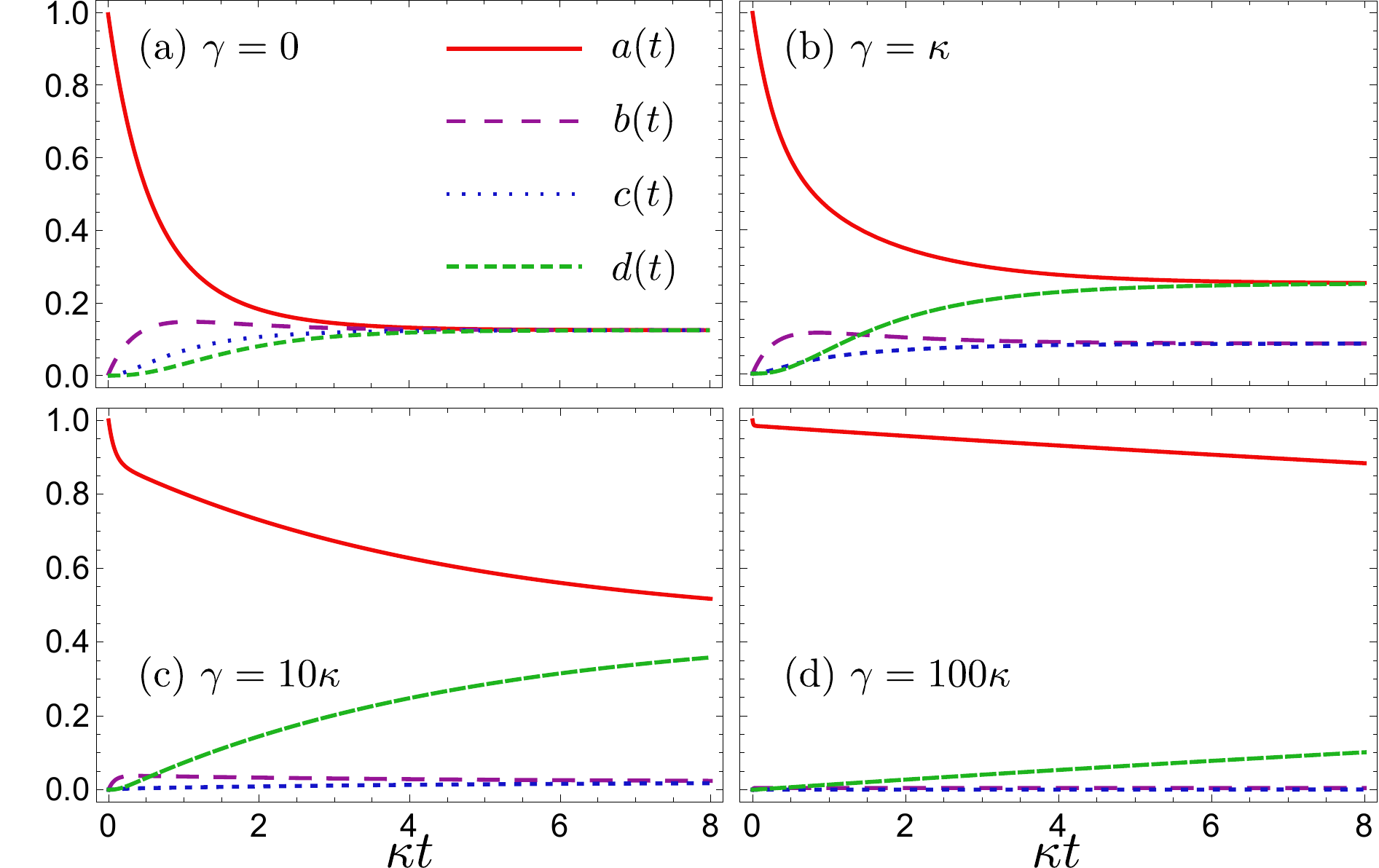}
	\caption{(Color online) Time-dependence of the coefficients in \eqref{eq:3q-ansatz} obtained from the solution of the master equation presented in Appendix~\ref{app:sol}, for the following values of the ratio $\gamma/\kappa$: $0$, $1$, $10$ and $100$.
\label{fig:coeff}}
\end{figure}

In Appendix~\ref{app:sol} we show that ansatz \eqref{eq:3q-ansatz} is correct and we derive analytical expressions for the coefficients $a(t)$, $b(t)$, $c(t)$ and $d(t)$, plotted in Fig.~\ref{fig:coeff} for several values of the ratio $\gamma/\kappa$. 
Without the continuous error correction, i.e. for $\gamma=0$,  the four coefficients converge to the value $1/8$ [see Fig.~\ref{fig:coeff}\,(a)].
When the error-correcting process is turned on [see Fig.~\ref{fig:coeff}\,(b) and (c)], the weight of $\rho(t)$ outside the code-space gets suppressed.
Finally, in the strong error-correction limit [see Fig.~\ref{fig:coeff}\,(d)], $b$ and $c$ are negligibly small at all times, while $a$ decreases very slowly and $d$, representing undetectable logical errors, increases at the same rate. 
Logical errors will eventually corrupt the information, but the CTQEC approach provides a method to make the storage time arbitrarily long by increasing $\gamma/\kappa$.

\subsection{Average recovery fidelity}

We now quantify the last statement and
compute the average fidelities \fid{\mc I} and \fid{\mc R} between the initially encoded qubit and its time-evolved counterpart, as defined in Eq.~\eqref{eq:average:fidelity1}.

Let
$\omega^L_{\mb n} = P \omega^L_{\mb n} P \in \states{  \mc H_{L} }$
be the encoded logical state corresponding to the unit vector $\mb n$ in the Bloch sphere, i.e. $\omega^L_{\mb n}=(P + \mb n\cdot \lpos )/2$, where $\lpos = (P \bar X , P\bar Y, P\bar Z)^T$. 
Accordingly the average fidelity \fid{\mc I}, i.e. the average fidelity associated with the CTQEC mapping~(\ref{dlog}) without any extra final recovery operation, is
\begin{eqnarray}
\fid{\mc I} &=& \int d\mu_{\mb n} \; \text{Tr} [ \, \omega_{\mb n}^L \mc D_t^{(\eta)} [ \omega_{\mb n}^L] \,] \label{eq:3q-ctqec-avg-fid}
 \\
%\end{equation} 
%(remember that here $\dec = \exp [ t(\mc L_{\text{e.c.}} +\mc L_{\text{noise}} ) ]$); recalling ansatz~\eqref{eq:3q-ansatz}, this becomes
%\begin{equation}
%\fid{\mc I}
&=& \int d\mu_{\mb n} \tr{ a(t)  \omega^L_{\mb n} +d(t)   \omega^L_{\mb n} \bar X   \omega^L_{\mb n} \bar X} \nonumber \\ 
 &=& a(t) +d(t) \hspace{-0.05cm} \int \hspace{-0.1cm} d\mu_{\mb n} \frac{1+n_x^2 -n_y^2 -n_z^2}{2} \nn 
 \hspace{-0.1cm} = a(t)+\frac{1}{3} d(t),
\end{eqnarray}
where we used the fact that  $PX_iP=0$ $\forall \, i$, and the fact that 
 $\bar X$ acts on the encoded qubit as a bit-flip operator.

On the contrary performing an additional QEC operation $\mc T=\mc R$ at read-out corresponds to the following update rule for the state coefficients:
\al{
a^\prime & = a+ 3b\,, & b^\prime & =0\,, &
c^\prime & = 0\,, & d^\prime & =d+3c\,.
}
The effect of this final, read-out operation on the recovery fidelity (\ref{eq:average:fidelity1}) is 
\begin{equation}
\fid{\mc R} = a(t)+3b(t) +c(t) +\frac{1}{3} d(t)  
 = \fid{\mc I} + 3b(t) +c(t)\,. \nonumber
% \label{eq:3q-ctqec-avg-fid-ub}
\end{equation}

Clearly $\fid{\mc R}  \geq \fid{\mc I}$, but in the strong error-correction regime
the relative difference is small, since coefficients $b$ and $c$ are much smaller than $a$ or $d$ at all times.

\begin{figure}
	\centering
	\includegraphics[width=\columnwidth]{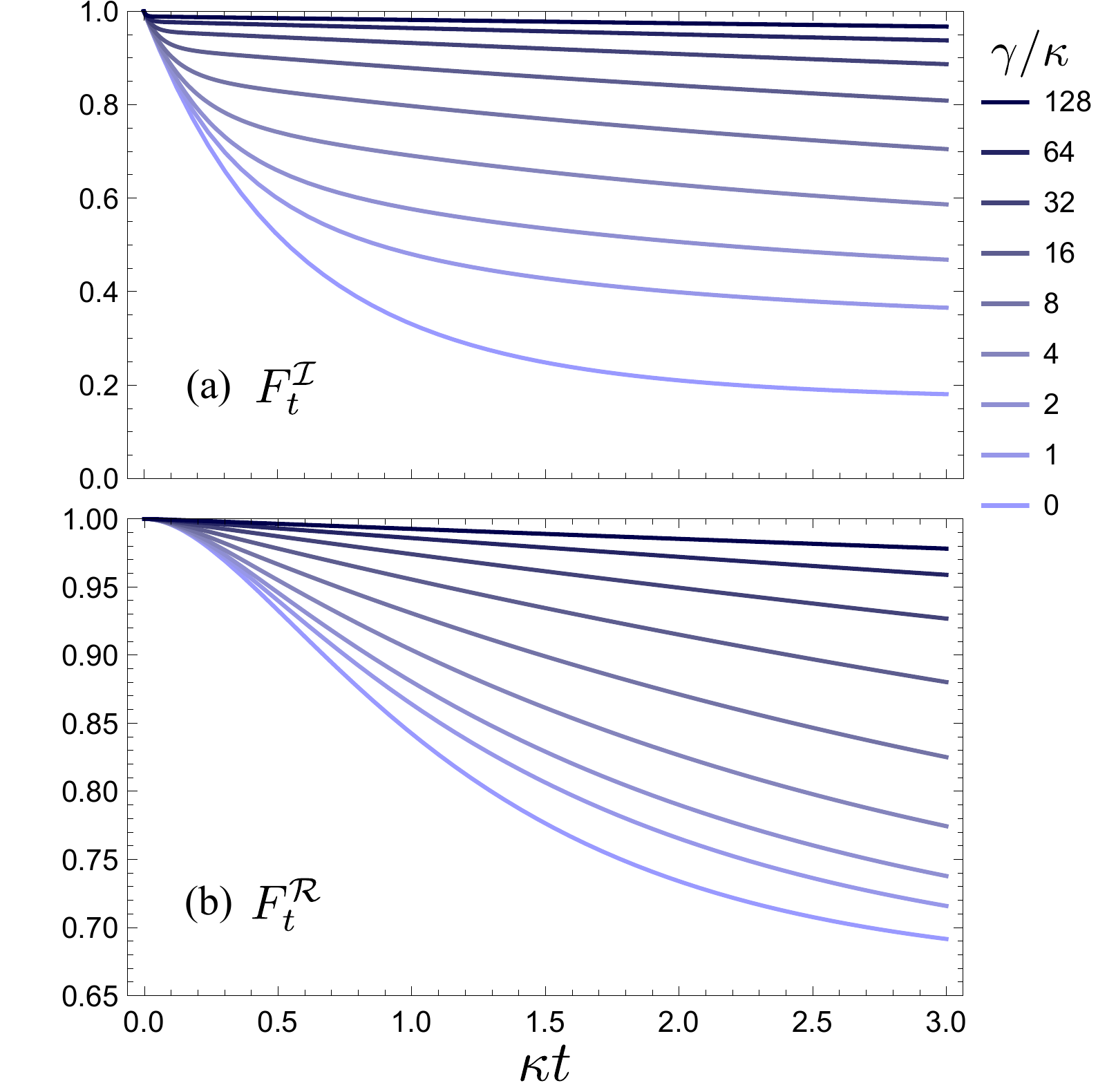}
	\caption{(Color online) Decay of the average recovery fidelities \fid{\mc I} and \fid{\mc R} from Eq.~(\ref{eq:average:fidelity1})
	 for the 3-qubit code exposed to bit-flip noise of strength $\kappa$ and continuous QEC of strength $\gamma$. (a) No additional recovery operation is applied at read-out. 
	(b) A final round of QEC is applied at read-out.
	\label{fig:3qf}}
\end{figure}

\begin{figure}
\centering
\includegraphics[width=0.9\columnwidth]{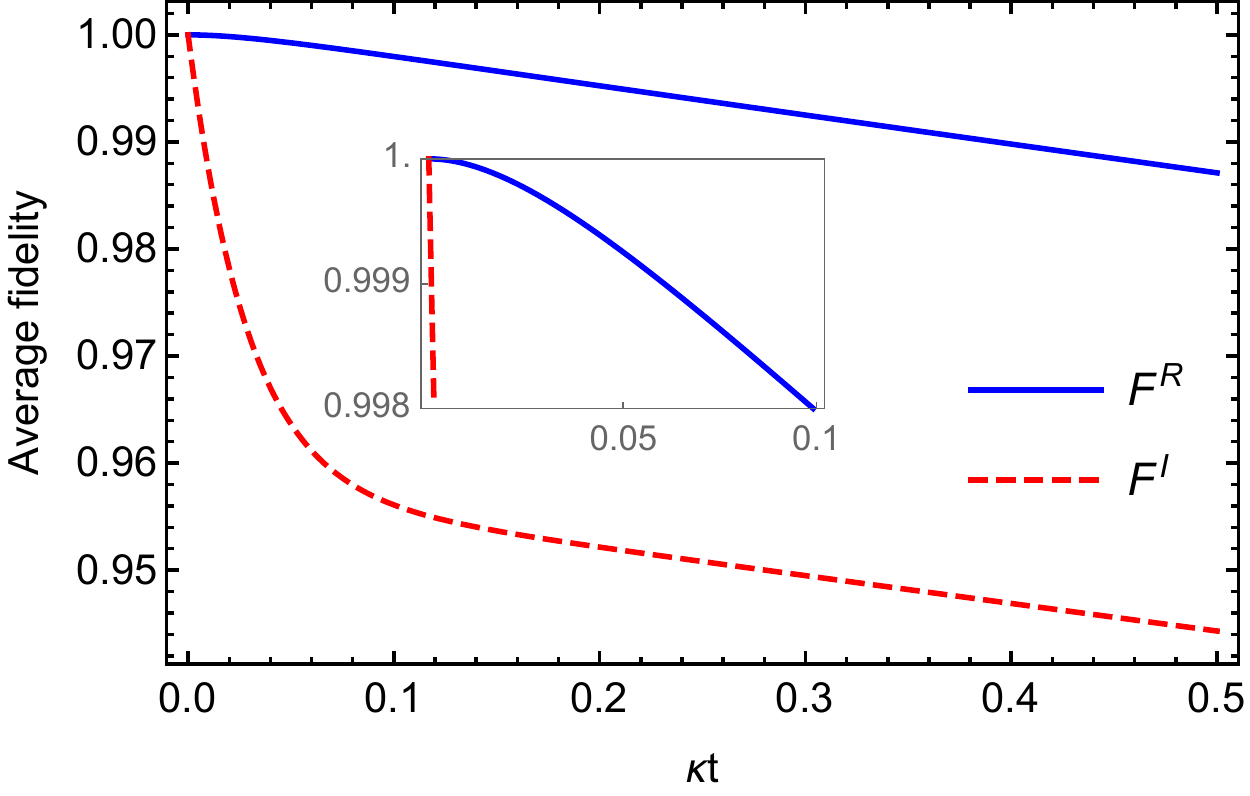}
\caption{(Color online) 
Comparison between the average recovery fidelities $\fid{\mc R}$ (dashed line) and $\fid{\mc I}$ (solid line) for  the 3-qubit code exposed to bit-flip noise of strength $\kappa$ and continuous QEC of strength $\gamma=32\kappa$.
The final round of QEC compensates for the fidelity loss in the initial $t \lesssim \frac{1}{\gamma}$ ($\kappa t \lesssim 0.03$) transient, but not for the subsequent slow decay. 
The inset shows the quadratic behavior of \fid{\mc R} at short times.
\label{fig:comp}}
\end{figure}

We are now in a position to first benchmark the results of Sec.~\ref{sec:perturb}. 
Since the fidelity is linear in $\mathcal D^{(\eta)}_t$ by  definition [see Eq.~\eqref{eq:average:fidelity1}],
Eq.~\eqref{uno1} and~\eqref{ff} predict that for $\gamma / \kappa \gg 1$ and at short times, one should have 
$\fid{\mc I} \sim 1 - \alpha \kappa t$ and $\fid{\mc R} \sim 1 - \beta \kappa^2 t^2$, where $\alpha$ and $\beta$ are real coefficients independent of $\gamma$. 
This behavior is displayed in Fig.~\ref{fig:3qf}, which moreover shows that it extends to all values of $\gamma / \kappa$ (i.e. not only in the strong correction regime).
The role of the final recovery operation is shown in detail in Fig.~\ref{fig:comp}:
an initial and abrupt fidelity loss $\sim \kappa t$ for $\fid{\mc I}$ is turned into a more gentle decay after time $1/\gamma$. The application of a final recovery operation eliminates also the first fast decay and lets a more gradual decay $\sim \kappa^2 t^2$ set in.
From an analytical standpoint, the expressions for the averaged fidelities in Appendix~\ref{app:sol}, Eq.~\eqref{eq:AppB:FI} and~\eqref{eq:AppB:FR}, can be expanded for short times and correctly reproduce the general properties derivable from Sec.~\ref{sec:perturb}: $\fid{\mc I } \sim {1 -\frac{3}{2} \kappa t}$, $\fid{\mc R} \sim 1-\frac{1}{2} \kappa^2 t^2$.

%----------------------------------------------------------------------------------------
% FIXED POINTS
%----------------------------------------------------------------------------------------

\subsection{Geometric picture\label{subsec:geom} }

Finally, taking advantage of the mathematical simplicity of the model, we derive explicitly the eigenmodes of the dynamics and the associated decay rates, which
remarkably can be expressed exactly for all values of $\kappa / \gamma$.
This exact analysis of the spectral properties of the dynamics provides an intuitive geometric interpretation to the general phenomenology that we discussed in the previous sections.

One quantity that is particularly interesting in light of the general perturbative picture of Sec.~\ref{sec:perturb} is the asymptotic decay rate (ADR)~\cite{adr} of the  Markovian evolution~(\ref{formalsol}). The latter is defined 
as the maximum $\lambda<0$ such that $\Re(\tilde{\lambda}) = \lambda$ with $\tilde{\lambda}$ being an eigenvalue of the Lindbladian generator  of the
system, i.e. 
\begin{eqnarray} \label{fixed} 
\mc L [\theta] =  \tilde{\lambda} \, \theta \;, \qquad 
\mc L = \eta \mc L_{\text{n.}}+ \gamma \mc L_{\text{e.c.}} \;,
\end{eqnarray}  
with $\theta$  the associated eigenoperator. 

The fixed point equation~(\ref{fixed}) 
can be  solved 
passing through  the Liouville representation~\citep{HOLEVOBOOK} where $\mc L$ is mapped into 
a $64\times 64$ matrix ${\mc M}_{\mc L}$ whose spectrum can be solved by casting it in the standard Jordan form (the matrix not being Hermitian in general). 
For the model we are considering we get 
\begin{align}
\mc M_{\mc L} 
 =& \left(\frac{\kappa}{2} I \otimes I + \gamma P \otimes P \right) \left( I \otimes I + \sum_{i=1}^3 X_i \otimes X_i \right) + \nonumber \\
 &-(2\kappa+\gamma) I \otimes I
 \label{eq:fixed-points-bigbadmatrix}
\end{align}
whose eigenvalues are listed in Table~\ref{tab:eigen}. 

In the strong QEC limit ($\gamma \gg \kappa$) there are
2 stable modes,
60 modes that decay over a short time scale $\sim~\gamma^{-1}$,
and 2 modes that decay over a longer time scale $ \sim~\gamma~\kappa^{-2}$:
for $\gamma \gg \kappa$, the latter reads:
\begin{equation}
-\frac{(\gamma+4\kappa) - \sqrt{(\gamma+4\kappa)^2-12\kappa^2} }{2} 
 = -\frac{3\kappa^2}{\gamma} + 
\mc O \left( {\kappa^3}\right)\,. 
\label{eq:adr}
\end{equation}
These modes are  stable in the limit of infinitely strong QEC (or, equivalently, in the absence of noise).
For large but finite values of $\gamma / \kappa$, they are the slowest decaying modes and define the ADR of the problem.

\subsubsection{Stable modes.}
The first stable mode is easily found by observing that the subspace spanned by  the operators $I$ and $P$ is invariant under the action of the Lindbladian  
$\mc L$. Indeed we have
\al{
\mc L [I] 
& = \gamma (4P-I)\,, &
\mc L[P] 
& = \frac{\kappa}{2} (I-4P)\,.
}
The first fixed point is thus
\al{
\approxlpo 0 & =
\frac{1}{2} \frac{\kappa I +2\gamma P}{2\kappa+\gamma}\,.
}
By the same reasoning, applied to operators $\bar X$ and $P \bar X P$, it can be shown that the second fixed point is
\al{
\approxlpo{1} & = \frac{1}{2} \frac{ \kappa \bar X+ 2 \gamma P \bar X P}{2\kappa+\gamma}\,.
}
The normalization is chosen so that $\tracenorm{\approxlpo 0} = \tracenorm{ \approxlpo 1}= 2$.
In the strong QEC limit, $\approxlpo{0}$ approximates $\lpo{0} \doteqdot P$ (the identity operator on the logical qubit) and $\approxlpo{1}$ approximates $\lpo{1} \doteqdot P\bar X$ (the Pauli operator $\sigma_1$ on the logical quibt).

\subsubsection{Slowly-decaying modes.}
The other two modes of interest are those that, though not fixed, decay very slowly in the $\gamma \gg \kappa$ limit. 
These two modes are expected to approximate $\lpo 2$ and $\lpo 3$.
It is easy to see that the subspace spanned by $\bar Z$ and $P \bar Z P$ is invariant under the action of the Lindbladian:
\eqsys{
& \mc L [\bar Z] 
=-(3\kappa+\gamma) \bar Z - 2 \gamma P \bar Z P\,, \\
& \mc L[P \bar Z P] 
=  -\frac{\kappa}{2} \bar Z - \kappa P \bar Z P\,.
}
The eigenvalue problem restricted to this subspace can be solved to obtain the eigenvalues
\al{
\lambda_\pm = \frac{-(\gamma+4\kappa) \pm  \sqrt{(\gamma+4\kappa)^2-12\kappa^2} }{2}\,.
}
$\lambda_+$ is the ADR (see also Table~\ref{tab:eigen}) and
the corresponding slowly-decaying mode is
\al{
\approxlpo 3 & =\frac{\left( 2\kappa+\gamma+\chi \right)P\bar Z P - \kappa \bar Z}{4\kappa+\gamma+\chi }\,,
}
where  we introduce the shorthand notation $\chi =  \sqrt{(\gamma+4\kappa)^2-12\kappa^2}$.
In the strong QEC limit this approximates $\lpo{3} \doteqdot P \bar Z$.
Finally, by applying the same reasoning to $\bar Y$ and $P \bar Y P$, we get the second slowly-decaying mode:
\al{
\approxlpo 2 & =\frac{\left( 2\kappa+\gamma+\chi \right)P\bar Y P - \bar Y}{4\kappa+\gamma+ \chi }\,,
}
which completes the set of independent ``slow'' operators.

% table
\begin{table}
\centering
\begin{tabular}{cc}
\hline\hline
eigenvalue & multiplicity \\ 
\hline
0 & 2 \\
$ -\frac{1}{2} \left(\gamma+4\kappa- \sqrt{(\gamma+4\kappa)^2-12\kappa^2} \right)$ & 2 \\
$-\gamma$ & 6 \\
$-(\gamma+\kappa)$ & 22 \\
$ -\frac{1}{2} \left(\gamma+4\kappa + \sqrt{(\gamma+4\kappa)^2-12\kappa^2} \right)$ & 2 \\
$-(\gamma+2\kappa)$ & 24 \\
$-(\gamma+3\kappa)$ & 6 \\
\hline\hline
\end{tabular}
\caption{Eigenvalues of the 3-qubit Lindbladian $\mc L$ of Eq.~(\ref{eq:3q-ctqec-mastereq-final})  involving both bit-flip noise and CTQEC.
\label{tab:eigen}}
\end{table} 
% end of table

\subsubsection{Qualitative behavior of the encoded qubit.}

Let us now show that the previous identification of stable and slowly-decaying modes allows us for a pictorial understanding of the real-time corruption of information which takes place under the action of
$\dec^{(\eta)} \circ \mathcal R$ (no recovery operation acting at the end).

\begin{figure}
\centering
\includegraphics[width=0.9\columnwidth]{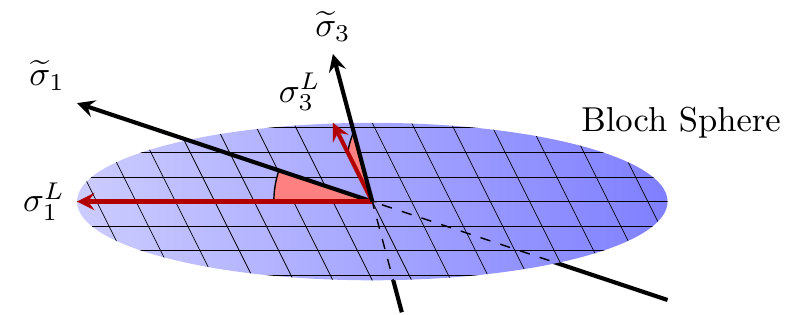}
\caption{(Color online) The logical Bloch sphere embedded in the larger state space. 
The logical operator $\lpo i$ is in general not parallel to the stable or slow-decaying operator $\approxlpo i$. 
The angle goes to zero in the limit of strong QEC (or weak noise). \label{fig:bloch}}
\end{figure}

The set of stable and ``slow'' operators we found defines a quasi-stable 3-manifold $\mc Q \subset \states{\mc H_3}$:
\al{
\mc Q & = \set{\frac{\approxlpo{0} + \mb r \cdot \approxlpos}{2}, \ \mb r \text{ such that } \approxlpo{0} + \mb r \cdot \approxlpos  \geq 0}\,.
}
$\mc Q$ is generally \textit{not} perfectly parallel to the encoded Bloch sphere
(it is if and only if $\gamma / \kappa \to \infty$). 
Even when $\gamma/\kappa$ is large but finite, there is a small ``tilt'' between the two 3-manifolds, which  causes an unavoidable loss of fidelity at short times.
The situation is illustrated in Fig.~\ref{fig:bloch}.

Indeed, the first part of the time evolution is a sudden collapse of the encoded Bloch sphere onto $\mc Q$, which causes a small fidelity loss of order $\kappa/{\gamma}$ (the angle between the two manifolds) in a short time interval $\sim \gamma^{-1}$. Thus the the initial slope of $\fid{\mathcal I}$ is approximately independent of $\gamma$. 
This constitutes an intuitive geometric frame for the behavior of the fidelity curves in Fig.~\ref{fig:3qf}, and more generally for the initial transient predicted for all implementations of CTQEC in Sec.~\ref{sec:perturb}.

After this transient, when all fast-decaying modes have been suppressed, the dynamics is confined to the slow sub-manifold $\mc Q$. 
The decoherence process takes the form of an effective bit-flip channel whose strength is the ADR, $\Delta \propto \kappa^2 \gamma^{-1}$. 

\subsubsection{Comparison with perturbative calculations \label{sec:pert-vs-exact} }

We are now in a position to benchmark against this set of exact properties one of the results of the perturbative calculations of Sec.~\ref{sec:perturb}, 
namely that
$\mc R \circ \mc D^{(\eta)}_t \circ \mc R \approx \mc R + \frac{\eta^2 t}{\gamma} \; \mc R \circ \mc L^2_{\text{n.}}   \circ \mc R $ [see Eq.~\eqref{due2}].
A convenient expression for this channel is found in
Eq.~\eqref{eq:2nd-order-stab}, which describes the action of the competing noise and correction processes as an effective map on the code-space spanned by $\{\lpo 0 , \dots \lpo 3\}$. 
It is easy to verify that for the code and noise model we are considering now, \eqref{eq:2nd-order-stab} reduces to
\begin{equation}
\mc R \circ \mc L_{\text{n.}}^2  \circ \mc R (\rho)
= \sum_{\substack{i, j=1 \\ j\neq i}}^3 (\bar X \rho \bar X - \rho)\,,\end{equation}
which, given the commutation relations of logical operators, is $0$ if $\rho = \lpo 0 $ or $ \lpo 1$ and $-12 \rho$ if $\rho = \lpo 2$ or $\lpo 3$.
Since in this case the noise strength is $\eta = \frac{\kappa}{2}$,  the approximate channel \eqref{EQNEW1222} takes the following form:
\begin{equation}
\mc R \circ \mc D_t^{(\eta)} [\lpo{i}] =
	\begin{cases}
	 \lpo{i} \text{ if } i = 0,1 \\
	 \left[1-3\frac{\kappa^2}{\gamma}\left(t-\frac{1-e^{-\gamma t} }{\gamma}\right) \right]\lpo{i} \text{ if } i =2,3.
	\end{cases}
\label{eq:effective-bitflip}
\end{equation}
This is a bit-flip on the logical qubit.
For times such that $\kappa^2 t /\gamma \ll 1$, this correctly reproduces the ADR in Eq.~\eqref{eq:adr}.
Even though the validity of the perturbative approach is necessarily limited in time, there still is a wide time frame ($1/\gamma \ll t \ll \gamma/\kappa^2$) in which the long-time behavior of the exact solution is correctly captured, at least for appropriately small values of $\kappa/\gamma$.

The most complete comparison between the exact solution and the perturbative one comes from the averaged fidelity \fid{\mc R}.
An exact formula for said fidelity is given in App.~\ref{app:sol}, Eq.~\eqref{eq:AppB:FR}, whereas for the effective bit-flip channel \eqref{eq:effective-bitflip} it is easily seen to be $\frac{2}{3}+\frac{\lambda(t)}{3}$, $\lambda(t)$ being the eigenvalue in square brackets in \eqref{eq:effective-bitflip}.

\begin{figure}
\centering
\includegraphics[width=0.9\columnwidth]{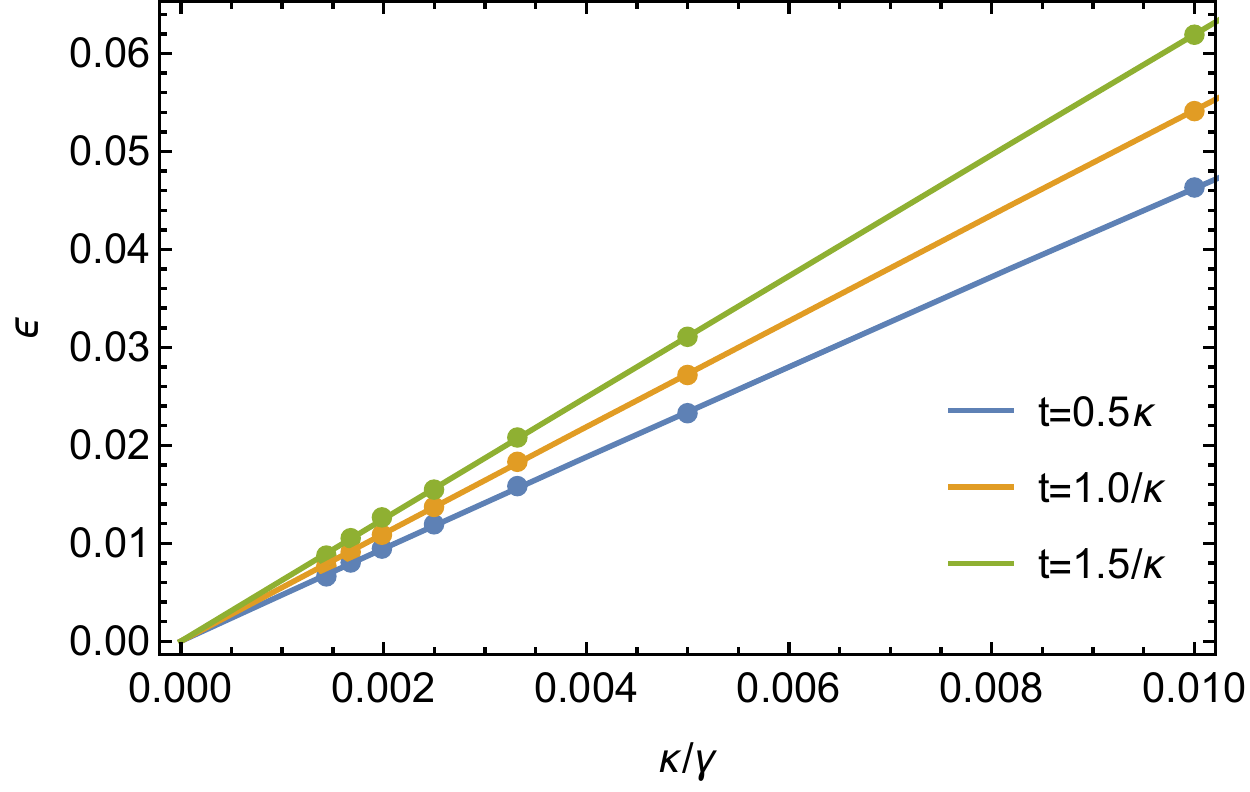}
\caption{
(Color online) Relative error of the approximation to the fidelity loss $1-\fid{\mc R}$ for the 3-qubit code discussed in Sec.~\ref{sec:pert-vs-exact}, $\epsilon = (1-\fid{\mc R})_{\text{pert.}}/(1-\fid{\mc R})_{\text{exact}}   - 1$.
The values are fitted very accurately by $\epsilon = 1.63\frac{\kappa}{\gamma} (\kappa t +2.34)$.
\label{fig:pert-vs-exact} }
\end{figure}

We compute the exact fidelity loss $(1-\fid{\mc R})_{\text{exact}}$ and its perturbative approximation $(1-\fid{\mc R})_{\text{pert.}}$ for several values of $t$ and of $\kappa/\gamma$ and plot the relative error of the approximation in Fig.~\ref{fig:pert-vs-exact}.
The relative error $\epsilon = {(1-\fid{\mc R})_{\text{pert.}} }/{(1-\fid{\mc R})_{\text{exact}}  } - 1$ is fitted very accurately by $\epsilon = 1.63\frac{\kappa}{\gamma} (\kappa t +2.34)$ in the relevant parameter range, thus proving that the correct behavior is reproduced in the strong-correction limit. More precisely,
an accurate approximation is obtained if $\kappa/\gamma \ll 1$ and $\kappa t \ll \gamma/\kappa$.
We note that the validity time is much longer than that guaranteed by the {\it a priori} perturbation theory bound, $kt \ll 1$.

%----------------------------------------------------------------------------------------
%----------------------------------------------------------------------------------------
% 5 QUBITS CTQEC
%----------------------------------------------------------------------------------------
%----------------------------------------------------------------------------------------

\section{CTQEC on the 5-Qubit Perfect Code}\label{sec:5q}

In this Section we test our perturbative approach on the smallest QEC code that corrects all single-qubit noise processes, which is the 5-qubit perfect code~\cite{laflamme}.
Since the dynamics of the model is not analytically solvable, we will resort to a numerical approach. The results are essentially similar to those derived analytically for the smaller 3-qubit code, and the more complex nature of this code gives us an opportunity to show how to practically apply the perturbative method in a more general setting.

\subsection{The 5-qubit code and the noise model}

The 5-qubit code is a stabilizer code defined by the 4 stabilizer operators
$\{g_n = X_n Z_{n \oplus 1} Z_{n \oplus 2} X_{n \oplus 3}:\, n=1,\dots 4\}$, ``$\oplus$'' denoting sum modulo 5. 
Measuring the 4 stabilizers yields $16$ possible syndromes: 
one is associated to the absence of any errors; 
the other 15 correspond to an $X$, $Y$ or $Z$ error occurring in any of the 5 qubits. 

The definition of the code-space projector reads $P = {\prod_{i=1}^4 (I+g_i)/2}$; the encoded logical operators are $\lpo{0} = P  $, $\lpo{1} = P \bar X $, $\lpo{2}=P \bar Y $, $\lpo{3}=P \bar Z $, where $\bar X = X_1 X_2 X_3 X_4 X_5$ and $[\bar X, P]=0$ (similarly for $\bar Y$ and $\bar Z$). 

The error-correcting Lindbladian~\eqref{eq:mastereq0} requires 16 Lindblad operators $\{ P U_{\mb s}\}$ and has again the intensity coefficient $\gamma$.
As for the noise model, we consider for simplicity a uniform depolarizing channel with strength $\kappa$, acting identically and independently on each qubit:
\begin{align}
 \mc N_t 
& = \phi_t^{\otimes 5}, & 
\phi_t \left( \frac{\sigma_0+ \mb a \cdot \bs \sigma}{2}\right) 
& = \frac{\sigma_0+ e^{-\kappa t}\mb a \cdot \bs \sigma}{2}\,.
\label{eq:5q-noise}
\end{align}
where the $\sigma_i$ are here the $2 \times 2$ identity and Pauli matrices.
The same discussion could be applied to other noise models with little algebraic differences.
The channel defined in Eq.~\eqref{eq:5q-noise}
is produced by a Markovian master equation whose Lindblad operators are  $\{ X_i, \, Y_i, Z_i\}$ and whose intensity coefficient is $\eta = \frac{\kappa}{4}$.

While for the 3-qubit QECC an analytical solution for the solution~(\ref{formalsol}) of  the dynamics
 was feasible,
the 5-qubit QECC requires either a numerical treatment or a perturbative approach in the spirit of Sec.~\ref{sec:perturb}. 

\subsection{Numerical approach \label{subsec:ctqec5-rec-fid}}

\subsubsection{Recovery fidelity}
\begin{figure}
\centering
\includegraphics[width=0.95\columnwidth]{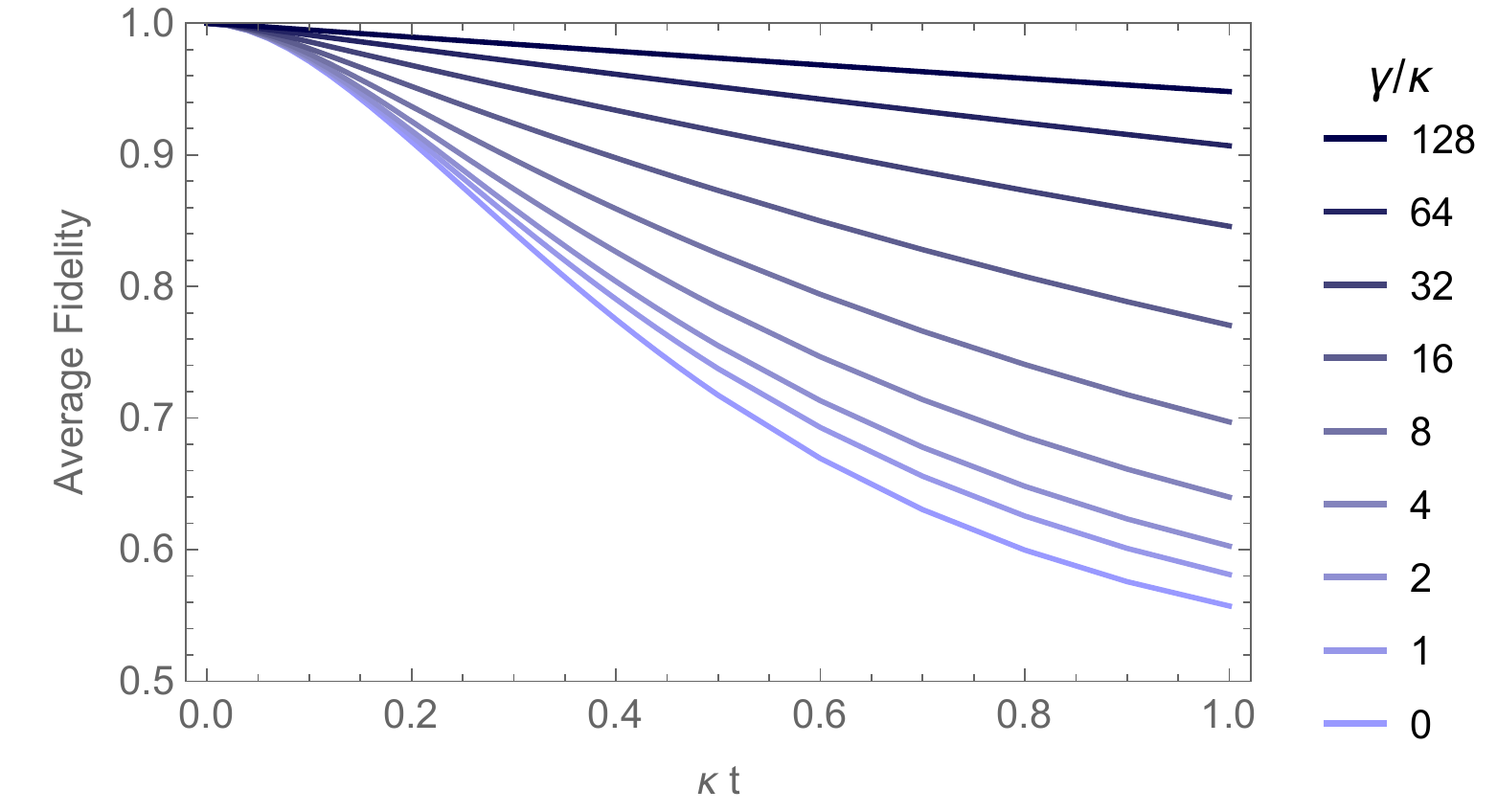}
\caption{(Color online) Decay of the average fidelity \fid{\mc R} for the 5-qubit perfect code subject to depolarizing noise of strength $\kappa$ and continuous error correction of strength $\gamma$, computed numerically for several values of the $\gamma/\kappa$ ratio.
\label{fig:5qf}}
\end{figure}

The numerical treatment benefits from the use of the Liouville representation, $\mc M_{\mc L}$ of the total Lindbladian $\mc L$
 which in this case  is given by
\begin{align}
	\mc M_{\mc L}
	=&  \left( \frac{\kappa}{4}I \otimes I + \gamma P \otimes P \right) \times \nonumber \\
	& \times \left( I \otimes I + \sum_{i=1}^5 (X_i \otimes X_i - Y_i \otimes Y_i + Z_i \otimes Z_i) \right) + \nonumber \\
 &-(4\kappa+\gamma) I \otimes I\,.
\end{align}
Analogously, we have for the recovery map $\mc M_{\mc R} = P \otimes P + \sum_{i=1}^5(PX_i \otimes P X_i  - PY_i \otimes P Y_i  + PZ_i \otimes P Z_i)$.
The average recovery fidelity from Eq.~\eqref{eq:3q-ctqec-avg-fid} (with an additional read-out operation $\mc R$) can be expressed in terms of these super-operators:  
\begin{align}
& \fid{\mc R}
= \int d\mu_{\mb n} \superbra{\omega_{\mb n}^L} \mc R \circ \mc D_t^{(\eta)} \superket{\omega_{\mb n}^L} \nn =\\
&\quad = \frac{1}{4} \superbra{\lpo 0} \mc M_{\mc R} e^{t \mc M_{\mc L}} \superket{\lpo 0} 
%\nn \\
%& \qquad  
+ \frac{1}{12} \sum_{i=1}^3 \superbra{\lpo i} \mc M_{\mc R} e^{t \mc M_{\mc L}} \superket{\lpo i}
\label{eq:5q-FRI}
\end{align}
As $\mc M_{\mc L}$ is a $1024 \times 1024$ matrix, the problem is numerically treatable.
In Fig.~\ref{fig:5qf} we plot 
the average fidelity \fid{\mc R} in Eq.~\eqref{eq:5q-FRI} for several values of $\gamma/\kappa$.

By comparing Fig.~\ref{fig:3qf} and \ref{fig:5qf}, it can be seen that the 5-qubit QECC behaves in the same qualitative way as the 3-qubit code discussed in Sec.~\ref{sec:3q}, in agreement with the general argument from Sec.~\ref{sec:perturb}.

\subsubsection{Spectral properties and geometric picture}

The situation can be described geometrically in terms of stable, quasi-stable and suppressed eigenmodes of the Lindbladian, analogously to the study in Sec.~\ref{sec:3q}.
Again, a numerical approach is necessary.

We choose several exponentially spaced values for $\gamma/\kappa$, namely $\gamma_n = 2^n \kappa$, $n \in \{0,1,\dots 12\}$; 
for each $\gamma_n$ we compute numerically the eigenvalues of $\mc M_{\mc L}$ and verify that its Jordan form is diagonal. 
The 1024 eigenvalues cluster into 8 sets of identical eigenvalues.
One is a non-degenerate $0$ eigenvalue (required by trace preservation); 
next, a threefold-degenerate ADR appears, scaling as $\gamma^{-1}$ for $\gamma \gg \kappa$; all the remaining eigenvalues scale as $\gamma$ 
The results are plotted in Fig.~\ref{fig:eigen}.

This behavior is analogous that of the 3-qubit code. The main difference is the absence of non-trivial fixed modes and the presence of 3, rather than 2, quasi-stable modes. This is however due entirely to the nature of the noise model: while the bit-flip noise considered in Sec.~\ref{sec:3q} preserves the $\hat x$ axis of the Bloch sphere, the depolarizing noise is completely isotropic, hence the 3-fold degeneracy.

\subsection{Perturbative approach} \label{sec:5qpert}

\begin{figure}
\centering
\includegraphics[width=0.9\columnwidth]{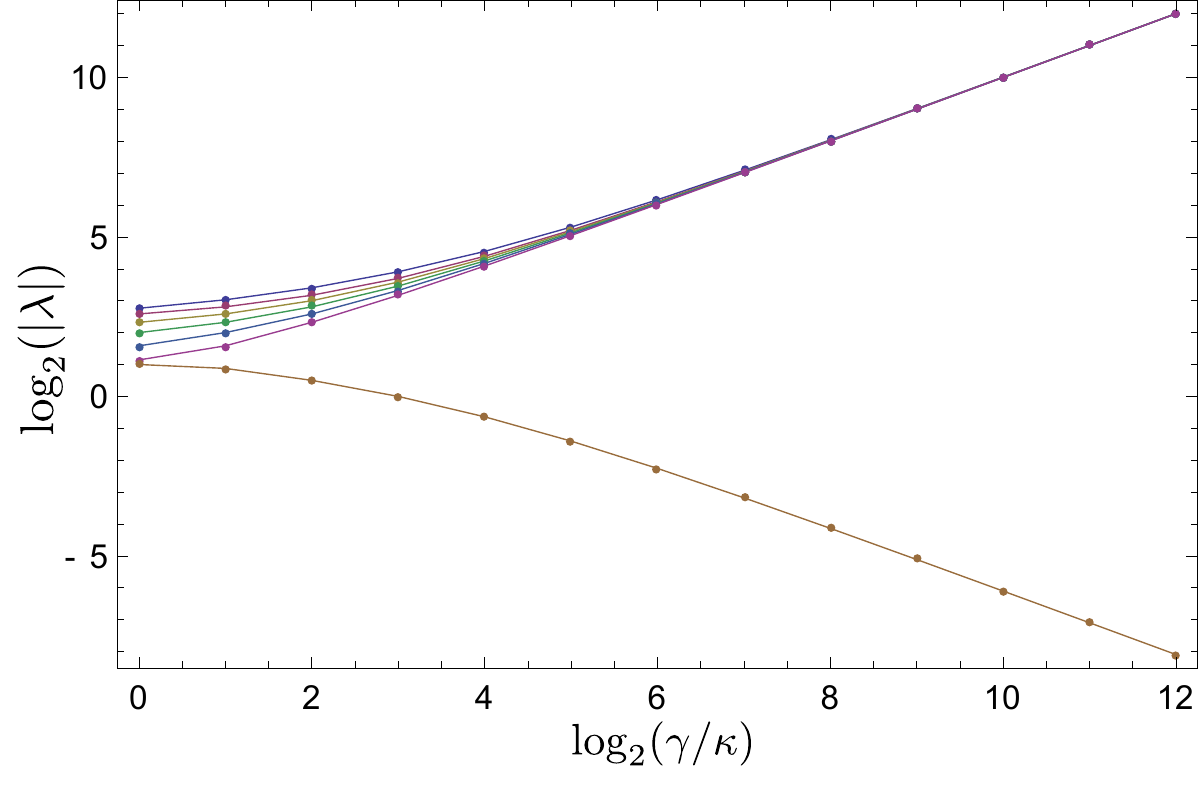}
\caption{(Color online) Absolute value of the eigenvalues $\lambda$ of $\mc L_{\rm e.c} + \mathcal L_{\rm noise}$, the Lindbladian superoperator for the 5-qubit code subject to depolarizing noise of strength $\kappa$ and continuous error-correction of strength $\gamma$. The 1024 eigenvalues cluster into 8 distinct values (the non-degenerate 0 eigenvalue is not shown). One of the eigenvalues drops $\sim \gamma^{-1}$, while the other 1020 increase $\sim \gamma^{+1}$. \label{fig:eigen} }
\end{figure}

We can now test the accuracy of the perturbative approach presented in Sec.~\ref{sec:perturb} against the numerical results.

The perturbative prediction can be obtained from Eq.~\eqref{eq:2nd-order-stab}.
In App.~\ref{app:5qpert} we prove that the approximate channel takes the form of an isotropic depolarizing noise on the logical qubit:
\begin{equation}
\mc R \circ \mc D_t^{(\eta)}  [\lpo{i}] =
	\begin{cases}
	 \lpo{0} \text{ if } i = 0 \\
	 \left[1-15\frac{\kappa^2}{\gamma}\left(t-\frac{1-e^{-\gamma t} }{\gamma}\right) \right]\lpo{i} \text{ if } i =1,2,3
	\end{cases}
\label{eq:effective-depol}
\end{equation}
Thus the perturbative estimate for the ADR is $15 \frac{\kappa^2}{\gamma}$, which is consistent with the numerical estimate one gets from the asymptote of the lower branch of Fig.~\ref{fig:eigen}.

\begin{figure}
\centering
\includegraphics[width=0.9\columnwidth]{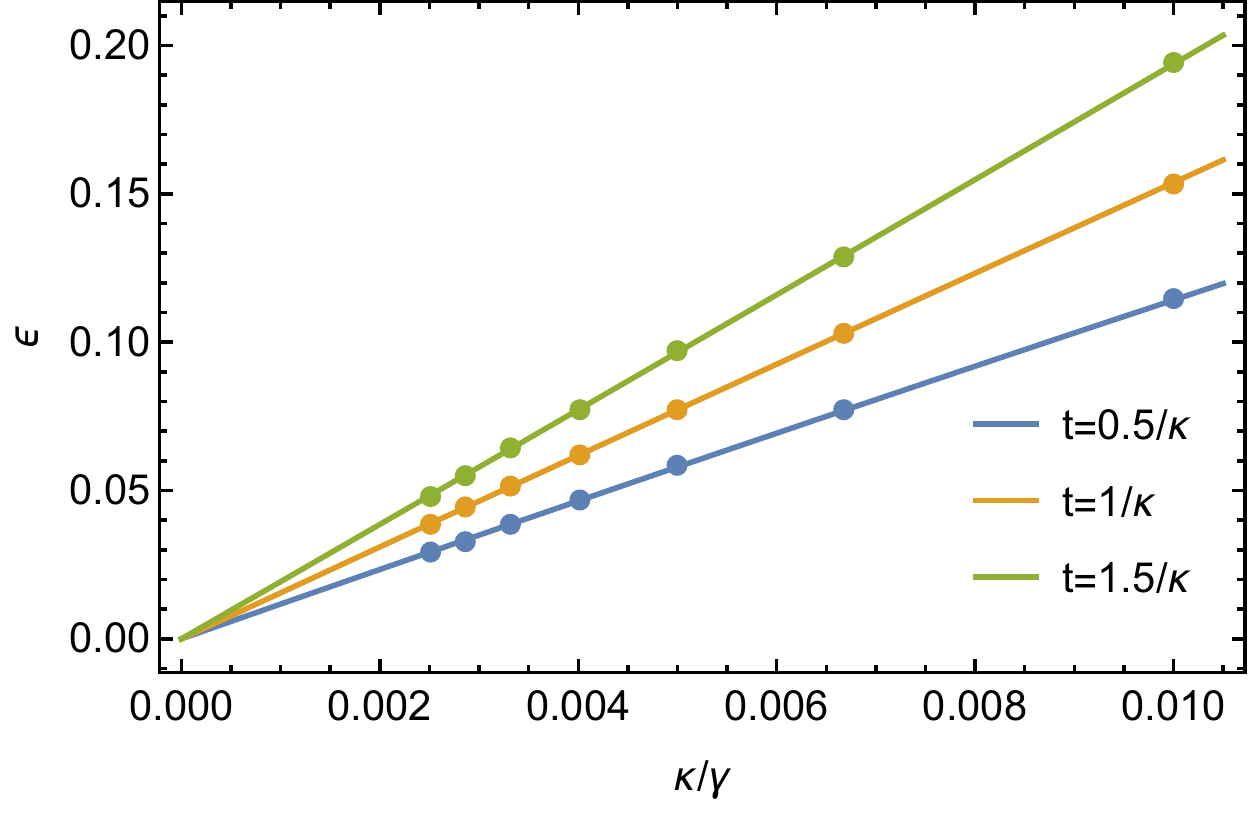}
\caption{
(Color online) Relative error $\epsilon$ of the approximation to the fidelity loss $1-\fid{\mc R}$ for the 5-qubit code discussed in Sec.~\ref{sec:5qpert}: 
$\epsilon = (1-\fid{\mc R})_{\text{pert.}}/(1-\fid{\mc R})_{\text{num.}}   - 1$.
We computed the relative error for several values of $\kappa/\gamma$ and $t$.
The values are fitted very well by $\epsilon = 7.9 \frac{\kappa}{\gamma} (\kappa t +0.94)$.
\label{fig:pert-vs-exact5q}}
\end{figure}

We turn again to the average fidelity \fid{\mc R} for a more thorough comparison.
The average fidelity for the effective depolarizing channel is $\frac{1+\lambda(t)}{2}$, $\lambda(t)$ being the expression in square brackets in \eqref{eq:effective-depol}.
This gives us the perturbative estimate of the fidelity loss, $(1-\fid{\mc R})_{\text{pert.}}$.
As a benchmark, we compute the fidelity loss numerically following the method described in the previous subsection, and label it $(1-\fid{\mc R})_{\text{num.}}$.
We then compare the two and plot the relative error $\epsilon = (1-\fid{\mc R})_{\text{pert.}}/(1-\fid{\mc R})_{\text{num.}}   - 1$ in Fig.~\ref{fig:pert-vs-exact5q}.
The relative error $\epsilon = (1-\fid{\mc R})_{\text{pert.}}/(1-\fid{\mc R})_{\text{exact}}   - 1$ is fitted very accurately by $\epsilon = 7.9 \frac{\kappa}{\gamma} (\kappa t +0.94)$ in the relevant parameter range.
An accurate approximation is obtained if $\kappa/\gamma \ll 1$ and $\kappa t \ll \gamma/\kappa$, in complete analogy to the 3-qubit case.
In this case, though, the fit coefficients are substantially larger [compare Fig.~\ref{fig:pert-vs-exact} and Fig.~\ref{fig:pert-vs-exact5q}].

\section{Conclusion\label{sec:conclusion} }

CTQEC can be interpreted both as a mathematical method for modeling discrete-time QEC and as a description of a suitably engineered dissipative process. 
It results in a competition between two different open-system Markovian dynamics, the noise and the error correction. 

We analyzed perturbatively the limit of strong QEC (or weak errors). By examining the bare noisy evolution, 
we derived a criterion for the effectiveness of recovery operations. 
By considering the second-order contribution for this class of ``effective'' recovery operations, we gained insight into the real-time dynamics of the encoded qubit.
We then compared these general predictions with some accurate results about the simplest instances of CTQEC, namely the continuous implementations of the 3-qubit bit-flip code and the 5-qubit perfect code. 
The main features of the time evolution predicted by our perturbative argument and shown by the examples are
a short transient in which QEC is ineffective, and a subsequent regime in which the decay rate for the encoded information is suppressed by a factor of $\kappa/\gamma$, $\gamma$ being the QEC rate and $\kappa$ being the error rate.

In order to be effective as quantum memories, such protocols require high QEC rates: $\gamma$ has to exceed $\kappa$ by several orders of magnitude. 
Remarkably, this is the very regime in which our perturbative approach becomes a reliable tool. 
We thus propose this method as an efficient way of predicting quantitative details of both  traditional, discrete-time QEC codes in the fast-operation limit and more general dissipation-based error-correction protocols.

In particular, as CTQEC is a special instance of 
the coding schemes discussed in Ref.~\cite{Wolf},
this work provides a first indication of the dynamical properties 
of the envisioned protection protocols.

\acknowledgments 
We thank Rosario Fazio and Alexander M\"uller-Hermes for comments and discussions.
This work was supported  by Regione Toscana POR FSE 2007-2013.
M. R. thanks Scuola Normale Superiore for hospitality.

\appendix

%=================================
% === PERTURBATIVE EXPANSION
%=================================

\section{Details on the perturbative expansion in the strong-QEC limit \label{app:perturb}}

In this appendix we show some technical details about the derivation of formula \eqref{EQNEW1221}, which is the key result of the second-order perturbative calculation discussed in Sec.~\ref{sec:perturb2nd}.

The first-order term in the Dyson series for $\exp [t (\gamma \mc L_{\text{e.c.}} +\eta \mc L_{\text{n.}})]$ is
\begin{equation}
\dec[\eta] \circ \mc R -\mc R
\simeq  \eta  \int_0^t dt_1 \;  \mc D_{t-t_1}^{(0)}  \circ \mc L_{\text{n.}} \circ  \mc R\,.
\end{equation}
Since $\dec[0] = e^{-\gamma t} \mc I + (1-e^{-\gamma t})\mc R$,
using the 1-effectiveness of $\mc R$ against $\mc N_t$ we can replace $  \mc D_{t-t_1}^{(0)}$ by the factor $e^{-\gamma(t-t_1)}$ and perform the integral, which yields
\begin{equation}
\dec[\eta] \circ \mc R -\mc R =  \frac{\eta}{\gamma}(1-e^{-\gamma t} ) \mc L_{\text{n.}} \circ  \mc R + O(\eta^2)\,. 
\end{equation}
The second order term in the Dyson series is
\begin{align}
\mc D_t^{(\eta)} & \circ \mc R  -\mc R = \text{(first order)} \nn \\ 
	& + {\eta^2}   \int_0^t dt_1 \int_0^{t_1} dt_2 \;  \mc D_{t-t_1}^{(0)}  \circ \mc L_{\text{n.}} \circ \mc D_{t_1-t_2}^{(0)}  \circ  \mc L_{\text{n.}}  \circ \mc R\,;
\end{align}
again $\mc D_{t_1-t_2}^{(0)}  $ can be replaced by $e^{-\gamma(t_1-t_2)}$ using the 1-effectiveness of $\mc R$, and the integral over $t_2$ yields ${\frac{1}{\gamma}(1-e^{-\gamma t_1})}$.
Therefore
\begin{widetext}
\begin{eqnarray}
\mc D_t^{(\eta)} \circ \mc R  -\mc R	& = & \text{(first order)} 
+ \frac{\eta^2}{\gamma} \int_0^t dt_1\;   \left\{ e^{-\gamma t}  (e^{\gamma t_1}  - 1)\, \mc I \: 
+(1-e^{-\gamma t_1}) \left(1-e^{-\gamma (t-t_1)} \right) \mc R  \right\} \circ \mc L_{\text{n.}}^2 \circ \mc R \;.
\end{eqnarray}
\end{widetext}
The integral over $t_1$ is now easily evaluated and yields the result presented in Eq.~\eqref{EQNEW1221}.

\section{Leading-order error process for $k$-effective recovery maps \label{app:keffect}}
In this Appendix we prove formulas \eqref{eq:k-eff-1} and \eqref{eq:k-eff-2}, which describe the leading-order error process for the general case of a $k$-effective recovery map $\mc R$, $k\geq 2$, in the case in which an additional application of $\mc R$ occurs at read-out.

The first non-zero term in the Dyson series for $\dec[\eta]$ is at order $\eta^{k+1}$. That is because at least $k+1$ copies of $\mc L_{\text{n.}}$ need to appear between the initial and final $\mc R$ maps in order to avoid cancellation (the initial $\mc R$ means that we are focusing on logic states). The leading correction is thus
\begin{align}
\mc R \circ \dec[\eta] \circ \mc R
& \approx  \mc R +\eta^{k+1} 
\mc R \circ  \mc L_{\text{n.}} \circ  \int_0^t dt_1 \cdots \int_0^{t_k} dt_{k+1} \nn \\
& \qquad \mc D_{t_1-t_2}^{(0)} \circ  \mc L_{\text{n.}} \circ  \cdots \circ \mc D_{t_k-t_{k+1}}^{(0)} \circ \mc L_{\text{n.}} \circ \mc R\;;
\end{align}
now, all instances of $\mc D_{t_i-t_{i+1}}^{(0)} $ can be replaced by {$c$-numbers} $e^{-\gamma (t_i-t_{i+1})}$. We are thus left with
\begin{align}
\mc R \circ \dec[\eta] \circ \mc R
& \approx \mc R + \eta^{k+1} \mc R \circ \mc L_{\text{n.}}^{k+1} \circ \mc R \nn \\
& \quad \times \int_0^t dt_1 \cdots \int_0^{t_k} dt_{k+1} e^{-\gamma(t_1-t_{k+1})}\;.
\end{align}
The integral over $t_{k+1}$ maps $e^{\gamma t_{k+1}}$ to $\frac{1}{\gamma}(e^{\gamma t_k}-1)$. Then the integral over $t_k$ maps this to $\frac{1}{\gamma^2}(e^{\gamma t_{k-1}} -1-\gamma t_{k-1})$. It is easy to see that each iteration adds a factor of $\frac{1}{\gamma}$ and subtracts a new term in the Taylor series expansion of the exponential. Therefore after $k$ integrals we are left with $\frac{\eta^{k+1}}{\gamma^k} \int_0^t dt_1 e^{-\gamma t_1} (e^{\gamma t_1} -\sum_{l=0}^{k-1} \frac{(\gamma t_1)^l}{l!})$.
Changing variable to $y=\gamma t_1$ yields the result presented in Eq.~\eqref{eq:k-eff-2}.

%=================================
% === 3 QUBIT SOLUTION
%=================================

\section{Exact solution of the 3-qubit CTQEC\label{app:sol} }

In this Appendix, we shall present some technical details on the exact solution of the CTQEC equations for the 3-qubit code.

Let us first of all introduce the following shorthand notation:
given a code-state $\rho(0)$, we define
$A=\rho(0)$ (no flips), $B = \sum_{i=1}^3 X_i \rho(0) X_i$ (1 flip), 
$C = \sum_{i=1}^3 X_i \bar X \rho(0) \bar X X_i$ (2 flips), 
$D=\bar X \rho(0) \bar X$ (3 flips).
The action of the total Lindbladian \eqref{eq:3q-ctqec-mastereq-final} on these operators is as follows:
\begin{equation}
	\left\{
	\begin{aligned}
	\mc L(A) & = \kappa B/2- 3\kappa A/2\,, \\
	\mc L(B) & = \left( 3\kappa/2+ 3 \gamma \right) A - \left( 3\kappa/2 + \gamma \right) B + \kappa C\,, \\
	\mc L(C) & = \left( 3\kappa/2 +3\gamma \right)  D - \left( 3\kappa/2 + \gamma \right) C + \kappa  B\,, \\
	\mc L(D) & = \kappa C/2-3\kappa D/2\,.
	\end{aligned}
		\right.
\end{equation}
This proves that the ansatz \eqref{eq:3q-ansatz} is correct:
the subspace spanned by $A$, $B$, $C$ and $ D$ is indeed invariant. 
The evolution equations for the coefficients $a(t)$, $b(t)$, $c(t)$ and $d(t)$ introduced in Eq.~\eqref{eq:3q-ansatz} read as follows:
\begin{equation}
	\pmat{ \dot a \\ \dot b  \\ \dot c  \\ \dot d }
	=
	\kappa
	\pmat{
	-\frac{3}{2}  & \frac{3}{2} + 3 \frac{\gamma}{\kappa} & 0 & 0 \\
	\frac{1}{2} & -\frac{3}{2} -3\frac{\gamma}{\kappa} & 1 & 0 \\
	 0 & 1 & -\frac{3}{2} -3\frac{\gamma}{\kappa} & \frac{1}{2} \\
	0 & 0 & \frac{3}{2} + 3 \frac{\gamma}{\kappa} & -\frac{3}{2}
	}
	\pmat{ a \\ b \\ c \\ d }
	\label{eq:3q-coeff-system}
\end{equation}
The initial conditions for all code-states are $a(0)=1$, $b(0)=c(0)=d(0)=0$.

Eq. \eqref{eq:3q-coeff-system} can be solved by expressing it in terms of the functions $f_\pm(t) \doteqdot a(t) \pm d(t)$ and $g_\pm (t) \doteqdot b(t) \pm c(t)$, which yields two decoupled systems.
The result is the following:
\al{
x(t) & = x_0 +x_1 e^{-\frac{\gamma+4\kappa-\chi}{2} t} +x_2 e^{-\frac{\gamma+4\kappa+\chi}{2}t} +x_3 e^{-(\gamma+2\kappa)t}  \,,
\label{eq:sol}
}
where $\chi=\sqrt{(\gamma+4\kappa)^2-12\kappa^2}$ and $x$ stands for either $a$, $b$, $c$ or $d$.
The coefficients $\set{x_i:\, x=a,b,c,d; \, i=0,1,2,3}$ are given in Table \ref{tab:coeff}.
\begin{table}
\centering
	\begin{tabular}{l|cccc}
		\hline \hline
		 $x_i$ & 0 & 1 & 2 & 3 \\
		\hline
		\\
		$a$ & $\dfrac{\kappa+2\gamma}{4(\gamma+2\kappa)}$ & $\dfrac{\chi+\gamma+\kappa}{4\chi} $ & 
		$\dfrac{\chi-\gamma-\kappa}{4\chi} $ & $\dfrac{3\kappa}{4(\gamma+2\kappa)} $ \\
		\\
		$b$ & $\dfrac{\kappa}{4(\gamma+2\kappa)}$ & $\dfrac{\kappa}{4\chi} $ & 
		$-\dfrac{\kappa}{4\chi}$ & $-\dfrac{\kappa}{4(\gamma+2\kappa)} $ \\
		\\
		$c$ &  $\dfrac{\kappa}{4(\gamma+2\kappa)}$ & $-\dfrac{\kappa}{4\chi} $ & 
		$\dfrac{\kappa}{4\chi}$ & $-\dfrac{\kappa}{4(\gamma+2\kappa)} $ \\
		\\
		$d$ & $\dfrac{\kappa+2\gamma}{4(\gamma+2\kappa)}$ & $-\dfrac{\chi+\gamma+\kappa}{4\chi} $ & 
		$-\dfrac{\chi-\gamma-\kappa}{4\chi} $ & $\dfrac{3\kappa}{4(\gamma+2\kappa)}$ \\
		\\
		\hline \hline
	\end{tabular}
\caption{Coefficients \set{x_i} that appear in Eq.~\eqref{eq:sol}. We set $\chi = \sqrt{\gamma^2+8\gamma \kappa+4\kappa^2}$. \label{tab:coeff}}
\end{table}
It follows that
\begin{multline}
\fid{\mc I}
= \frac{\kappa+2\gamma}{3(\gamma+2\kappa)} +\frac{\gamma+\kappa+\chi}{6\chi}e^{-\frac{\gamma+4\kappa-\chi}{2}t} \\
+\frac{\chi-\gamma-\kappa}{6\chi} e^{-\frac{\gamma+4\kappa+\chi}{2}t}+\frac{\kappa  }{\gamma+2\kappa} e^{-(\gamma+2\kappa)t} \,,
\label{eq:AppB:FI}
\end{multline}
while
\begin{multline}
\fid{\mc R}
= \frac{2}{3} +\frac{\gamma+4\kappa+\chi}{6\chi}e^{-\frac{\gamma+4\kappa-\chi}{2}t}\\
-\frac{\gamma+4\kappa-\chi}{6\chi} e^{-\frac{\gamma+4\kappa+\chi}{2}t}\,.
% \label{eq:fid-r-exact}
\label{eq:AppB:FR}
\end{multline}

%=================================
% === PERTURBATIVE CALCULATION ON 5 QUBITS
%=================================

\section{Application of the perturbative calculation to the 5-qubit code}
\label{app:5qpert}

In this Appendix we detail the perturbative calculation mentioned
in Sec.~\ref{sec:5qpert}.

In order to evaluate \eqref{eq:2nd-order-stab} for the 5-qubit code, it is convenient to recall that the logical operators $\lpo{\alpha}$ can be cast in the form $P \bar{\sigma}_\alpha$, where $\sigma$ stands for either $I$, $X$, $Y$ or $Z$ and the bar denotes action on all five real qubits: $\bar X = X_1 X_2 X_3 X_4 X_5$, etc. Since both the $U_{\mb s}$ and the the $\bar \sigma$ are Pauli operators, they either commute or anti-commute with one another; 
this relation can be conveniently encoded by a binary function $f(\sigma, \mb s)$ such that $U_{\mb s} \bar \sigma = (-1)^{f(\sigma,\mb s)} \bar \sigma U_{\mb s}$.
As an example, Table~\ref{tab:app:5q:U} provides the values of $f$ for $\sigma=X$.

\begin{table}
\centering
\begin{tabular}{c c cc c|c c c cc}
\hline\hline
$\bf s$ && $U_{\bf s}$ & $f( X,\bf s)$ &&& $\bf s$ && $U_{\bf s}$ & $f( X,\bf s)$ \\ 
\hline
0000 && $I$ & 0 &&& 1000 && $X_2$ & 0 \\
0001 && $X_1$ & 0 &&& 1001 && $Z_4$ & 1 \\
0010 && $Z_3$ & 1 &&& 1010 && $Z_1$ & 1 \\
0011 && $X_5$ & 0 &&& 1011 && $Y_1$ & 1 \\
0100 && $Z_5$ & 1 &&& 1100 && $X_3$ & 0 \\
0101 && $Z_2$ & 1 &&& 1101 && $Y_2$ & 1 \\
0110 && $X_4$ & 0 &&& 1110 && $Y_3$ & 1 \\
0111 && $Y_5$ & 1 &&& 1111 && $Y_4$ & 1 \\
\hline\hline
\end{tabular}
\caption{Correcting unitaries $U_{\bf s}$ and the function $f(X, \mathbf s)$ defined in Appendix~\ref{app:5qpert} for the 5-qubit code.
\label{tab:app:5q:U}}
\end{table} 

Using these commutation relations, along with the fact that $U_{\mb r \oplus \mb s} U_{\mb r} U_{\mb s}$ has by definition trivial syndrome and hence commutes with $P$, one can rewrite \eqref{eq:2nd-order-stab} as
\begin{multline}
\mc R \circ \mc L_{\text{n.}} \circ \mc L_{\text{n.}} \circ \mc R (P \bar \sigma)
= - P \sum_{\mb r, \mb s} (\bar \sigma - U_{\mb r \oplus \mb s} U_{\mb r} U_{\mb s} \bar \sigma U_{\mb s} U_{\mb r}   U_{\mb r \oplus \mb s}) \\
= - \sum_{\mb r, \mb s} \left( 1-(-1)^{f(\sigma,\mb s) +f(\sigma,\mb r)+f(\sigma,\mb s \oplus \mb r) }\right) P \bar \sigma \,.
\end{multline}
The effective channel is thus diagonal in the basis of logical operators, with eigenvalues determined by the sum 
\begin{equation}
S(\sigma) = \sum_{\bf s \neq 0} \sum_{\bf r \neq 0} \left( 1-
 (-1)^{f (\sigma, \mathbf r)+f (\sigma, \mathbf s)+f (\sigma, \mathbf {r+s})} \right)\,.
 \end{equation}
This sum can be evaluated analytically for $ \sigma = I$: 
since $f (I, \mathbf s) = 0$ $\forall \mathbf s$, we simply have $S(I) = 0$.
All the other choices of $\sigma$ are equivalent, being related by a permutation of the syndrome indeces $\mathbf s$. The evaluation yields $S(X) = 240$, 
so that plugging in the noise strength $\eta = \frac{\kappa}{4}$ we get a three-fold degenerate eigenvalue of $-15 \frac{\kappa^2}{\gamma} \left( t-\frac{1-e^{-\gamma t}}{\gamma} \right)$, relative to the $\{\lpo{1},\lpo{2},\lpo{3}\}$ subspace, and a 0 eigenvalue relative to $\lpo{0}$.
This leads to the expression given in Eq.~\eqref{eq:effective-depol}.


\begin{thebibliography}{99}
\bibitem{NOTA}  It is worth stressing that the idempotent property~(\ref{idem}) plays a fundamental role in validating the correctness of the 
 perturbative expansion of $\mc D_t^{(\eta)}$ by forcing  $\mc R$ (resp. $\mc L_{\text{e.c.}}$) to exhibit a gapped spectrum formed by the eigenvalues 1 and 0
 (resp. 0 and -1).
\end{thebibliography}
\end{document}